\documentclass[utf8]{FrontiersinHarvard} 

\usepackage{url,hyperref,lineno,microtype}
\usepackage[onehalfspacing]{setspace}

\def\keyFont{\fontsize{8}{11}\helveticabold }
\def\firstAuthorLast{Miloshevich {et~al.}} 
\def\Authors{George Miloshevich\,$^{1,2}$, Philippine Rouby-Poizat\,$^{2}$, Francesco Ragone\,$^{3,4}$ and Freddy Bouchet\,$^{2,5}$}
\def\Address{$^{1}$Laboratoire des Sciences du Climat et de l’Environnement, UMR 8212 CEA-CNRS-UVSQ, Université Paris-Saclay \& IPSL, CE Saclay l’Orme des Merisiers, 91191 Gif-sur-Yvette, France \\
$^{2}$ENSL, CNRS, Laboratoire de Physique, F-69342 Lyon, France \\
$^{3}$Royal Meteorological Institute of Belgium, Brussels, Belgium \\
$^{4}$Université catholique de Louvain, Louvain-la-Neuve, Belgium \\
$^{5}$LMD/IPSL, ENS, Université PSL, École Polytechnique CNRS, France}
\def\corrAuthor{Corresponding Author}

\def\corrEmail{gmiloshe@lsce.ipsl.fr}

\begin{document}
\onecolumn
\firstpage{1}

\title[Robust heatwave teleconnections]{Robust intra-model teleconnection patterns for extreme heatwaves} 

\author[\firstAuthorLast ]{\Authors} 
\address{} 
\correspondance{} 

\extraAuth{}

\maketitle

\begin{abstract}

We investigate the statistics and dynamics of extreme heat waves over different areas of Europe. We find heatwaves over France and Scandinavia to be associated with recurrent wavenumber three teleconnection patterns in surface temperature and mid-tropospheric geopotential height. For heatwaves with return times of 4 years these teleconnection patterns and their dynamics are robustly represented in a hierarchy of models of different complexity and in reanalysis data. For longer return times, reanalysis records are too short to give statistically significant results, while models confirm the relevance of these large scale patterns for the most extreme heatwaves. A time series analysis shows that heatwave indices defined at synoptic scale are fairly well described by Gaussian stochastic processes, and that these Gaussian processes reproduce well return time plots even for very rare events. These results suggest that extreme heatwaves over different areas of Europe show recurrent typical behaviours in terms of long-range spatial correlations and subseasonal-scale temporal correlations. These properties are consistently represented among models of different complexity and observations, thus suggesting their relevance for a better understanding of the drivers and causes of the occurrence of extreme midlatitude heatwaves and their predictability.

\tiny
 \keyFont{ \section{Keywords:} heatwave, extreme, climate, reanalysis, teleconnection, return time, Ornstein-Uhlenbeck, autocovariance} 
 
\end{abstract}

\section{Introduction}


{Some of the most severe impacts of climate change are caused by rare and extreme events. Increased frequency and magnitude of extreme heatwaves is one of the most immediate and significant effects of global warming~\citet{van_oldenborgh_attributing_2022}. In the last decades a number of record breaking heatwaves have been observed \citet{ipcc_2014,Seneviratne21}. In the Northern Hemisphere mid-latitudes examples include the Western European heatwave of summer 2003, with a death toll of about 70, 000 \citet{Herrera2010}, the mega-heatwave over Russia of summer 2010 \citet{Otto_2012} with the deathtoll of about 100,000, and more recently the heatwave 2021 Western North America heatwave \citet{esd-2021-90}.} 

Recent studies have identified increased frequency of heatwaves occurring in mid-latitudes in Europe of duration longer than 6 consecutive days~\citet{rousi22}. In fact, the reasons for the high impact of the 2003 Western European heatwave was not only the magnitude of the temperature anomalies, but also its long duration (two successive heat events along an overall period of one month). In the case of the 2010 Russian heatwave, there was the compounding effect of high temperature, long duration (one month), and related wildfires. 2021 Western North America heatwave lasted about 10 days with peculiarly strong peak on a 2-day scale. In this case the main cause of the impacts where short term extreme temperature fluctuations, that were unprecedented in historical record~\citet{WWA21,schiermeier_climate_2021}. 

{Heatwaves typically occur when a stationary high pressure anomaly over a region leads to subsidence and increased incoming shortwave radiation fluxes, that increase surface and near surface temperatures ( \citet{horton16,PERKINS2015}). Short-term events lasting a few days are typically understood as caused by the occurrence of an atmospheric blocking over the region impacted by the heatwave. The most extreme and persistent events, on the other hand, require the activation of more particular dynamics and of feedback processes that enhance the surface temperature response and the duration of the event. This involves processes acting on space and time scales that go beyond the regional spatial scales and the synoptic time scales characteristic of standard heatwaves. A comprehensive analysis of the properties of extreme midlatitude heatwaves however is hindered by their rarity, the inherent noisiness of midlatitude weather fluctuations, and thus the difficulty to sample a sufficient number of events to provide a reliable statistics.}


{From a spatial point of view, it has been observed in the literature in the past years that some of the most extreme and persistent heatwaves seem to be associated with large scale atmospheric teleconnection patterns over the entire Northern hemisphere, for example  extreme heatwaves over the central USA \citet{Teng13}, Alberta \citet{petoukhov18}, and Western Europe \citet{Kornhuber_2019}. Following a recent classification of compound extreme events \citet{Zscheischler20}, this would correspond to spatially compounding events. Some authors have interpreted the presence of these patterns in terms of amplification of quasi-stationary Rossby waves, which are claimed to be related to the genesis of extreme events in different regions of the world during the same season \citet{kim10}; \citet{Petoukhov2013,petoukhov16}; \citet{Schubert11}.

The research in this area has been directed towards comprehending the pivotal role of waves that exhibit a wavenumber ranging from 5 to 8. To this end, an array of diverse detection techniques has been utilized encompassing empirical orthogonal functions (EOFs)~\citet{Teng13}, spectral analysis \citet{Kornhuber_2019,petoukhov18}, and indicators that rely on resonance models \citet{Petoukhov2013,petoukhov16}. Despite these strides, the precise identification of these teleconnection patterns is hindered by the limited amount of events available.} Furthermore, we are increasingly facing the questions ranging from estimating risks for similar ``black swan`` events found in the  tails of the distribution: from estimating their return times to attributing them to climate change~\citet{philip_protocol_2020}. This highlights the need to devise more efficient tools for extreme return time estimation.

{To overcome the sampling issue, \citet{Ragone18,Ragone:2020vs,Ragone21} have adopted rare event algorithm applied to simulations with climate models of different complexity. In this way they found that the warmest summers and most persistent heatwaves are robustly associated with teleconnection patterns with wavenumber 3-4, and were able to show that these patterns are indeed statistically significant (see in particular \citet{Ragone21}). However, a detailed analysis of the dynamical properties of these peculiar atmospheric states, and the possible relation with the amplification of higher wavenumber Rossby waves suggested in the literature, have not been investigated yet.}

{From a temporal point of view, there are two factors that characterise the most extreme and persistent events, up to the seasonal scale. First, these events typically occur when a train of heatwaves leads to an extreme hot summer, as in western Europe in 2003 \citet{cassou05}. This would correspond to temporal compounding mechanisms in \citet{Zscheischler20}. Possible reasons why a given season could be more likely to develop multiple heatwaves are not clear, but they would be of great interest for what concerns the seasonal to decadal predictability of temperature extremes. Second, it is  known that pre-existing anomalously dry soil conditions can amplify the intensity of heatwaves occurring at a later time, a typical mechanism for example of certain classes of heatwaves in Europe \cite{Stefanon_2012}. This would correspond to preconditioning mechanisms in \citet{Zscheischler20}. Preconditioning is typically studied with ad hoc numerical experiments initialized with dry or moist soil initial conditions, or in terms of correlations. A more detailed analysis of the properties of the decorrelation function of regional surface temperatures over different time scales seem to be absent from the literature.}

The goal of this paper is to better understand the properties of long range correlations in space and time characterizing extreme heatwaves, and to compare their robustness among models of different complexity and reanalysis data. We adopt a very simple definition of a heatwave event that specifically aims at the properties of time persistence of the temperature anomalies from the subseasonal to seasonal scale {(S2S)}. We analyse events over France and Scandinavia. In Section 2 we describe the data, the heatwave index, and the statistical tests and analysis adopted in this work. In Section 3 we describe the teleconnection patterns we find associated with the extreme events, we show the properties of the atmospheric activity associated with these events in terms of Hayashi spectra, and we discuss the statistical significance of the patterns. In Section 4 we analyse the temporal evolution of the events and the properties of the time autocovariance function of the regional temperatures, comparing them against a first order autoregressive process. Finally in Section 5 we discuss our conclusions.

\section{Methodology}\label{methodology}


\subsection{{Data}}

We  study extremely rare long lasting heatwaves over different areas of Europe during the June-July-August period (JJA). The data consists of daily averages of 2 meters temperature and 500 hPa geopotential height, from three different datasets: one intermediate complexity climate model (PlaSim), one CMIP5-type Earth system model (CESM 1.2), and the ERA5 reanalysis. 


\subsubsection{PlaSim}

The Planet Simulator \citet{fraedrich_2005,fraedrich_1998}, also known as PlaSim, is a climate model of intermediate complexity that has been developed to simulate the Earth's climate. The model's dynamical core uses a spectral transform method to solve the primitive equations governing vorticity, divergence, temperature, and surface pressure. The model has a horizontal resolution of T42 in spectral space, which translates to a spatial resolution of $2.8$ degrees by $2.8$ degrees or to $64\times128$ grid-points. Additionally, the model includes 10 vertical layers and incorporates simple parameterizations of the most important physical processes that influence the climate system, such as large-scale precipitation, clouds, moist and dry convection, boundary layer fluxes of latent and sensible heat, and vertical and horizontal diffusion and radiation. The atmospheric model is coupled to a simple bucket land surface scheme. To produce a stationary state climate close to that of the 1990s, the model is driven by prescribed values of sea surface temperatures, sea ice cover, greenhouse gas concentration, and incoming solar radiation fluxes at the top of the atmosphere. We used the model to generate 1000 years of data in this stationary state, this dataset has also been utilized in previous studies conducted by~\citet{Ragone18} and~\citet{jacques-dumas22}.


\subsubsection{CESM}
We use control runs from~\citet{Ragone21} which were performed with version 1.2.2 of the Community Earth System Model (CESM) developed by~\citet{CESM2013} utilizing an atmosphere and land only setup, with the active components being version 4 of the Community Atmospheric Model (CAM4) and version 2 of the Community Land Model (CLM2). The model was run in a statistically stationary state, with the concentration of greenhouse gases, sea surface temperature (SST), and sea ice cover held at values corresponding to the climate of year 2000.  The model features a horizontal resolution of $0.9$ and $1.25$ degrees in latitude and longitude, respectively, and includes 26 vertical layers in hybrid pressure coordinates.


\subsubsection{ERA5 reanalysis}\label{era5reanalysis}

We consider data from the ERA5 reanalysis dataset produced by the European Centre for Medium-Range Weather Forecasts (ECMWF) ~\citet{ecmwf2020,ecmwf2021}, from 1950 to 2020. 
The data was downloaded from \url{https://www.ecmwf.int/en/forecasts/datasets/reanalysis-datasets/era5} and re-gridded to a lower resolution of $241$ by $480$ while performing daily averaging over available 3 hour periods. 



\subsection{Definition of heatwaves and return times}\label{sec:heatwaves_def_returns}

To define heatwaves, several indices have been used in the literature, for different purposes \citet{PERKINS2015}. 
Many meteorological criteria used in  climate studies of temperature extremes  focus on sub-daily fluctuations (see for instance~\citet{ipcc_2014}). However, long-lasting heatwaves are the most detrimental to health and biodiversity \citet{barriopedro2011hot}. Moreover, many of the extreme heatwaves with the largest impact, for instance the Western European one in 2003 or the Russian one in 2010 lasted long, from two to five weeks.
These long lasting heatwaves were often composed of several sub-events compatible with the classical, short-time based definitions \citet{PERKINS2015}. The lack of comprehensive studies of the statistics of long-lasting events has actually been stressed in the last IPCC report \citet{Seneviratne21}. Moreover, many definitions that actually involve a measure related to the persistence of anomalous daily maximum temperature values with prescribed amplitude do not always carry a natural definition of a heatwave amplitude \citet{PERKINS2015}. This prevents to study independently impact of amplitude and duration of the heatwave and calls for a complementary definition of heatwaves, that can quantify both their amplitude in terms of temperature and their duration, in an independent way.

In this study we choose a specific criterion for selecting heatwaves (following the definitions of \citet{Ragone18,G_lfi_2019,Ragone:2020vs,Galfi21,Ragone21,galfi2021applications}), which consists of extremes of time averaged surface temperature fluctuations (anomalies)  defined as 
\begin{equation}\label{timeaveraged}
    A(t):=\frac{1}{T}\int_{t}^{t+T}\frac{1}{\mathcal{\left|D\right|}}\int_{D}\left(T_{s}-\mathbb{E}\left(T_{s}\right)\right)(\vec{r},u)\,\mathrm{d}\vec{r}\mathrm{d}u
\end{equation}
where $\mathbb{E}\left(T_{s}\right)(\vec{r},t)$ is the mean surface temperature at each point of the grid, with a seasonal and spatial variation. In the following the symbol $\mathbb{E}$ refers to the statistical empirical average over all available years, which corresponds to the duration of the datasets (1000 years for PlaSim and CESM and 71 years for ERA5. In this study the heatwave area $\mathcal{D}$  correspond either to France or Scandinavia and is depicted via shaded boxes on Figure~\ref{fig:local_std}. France is defined as a land enclosed between 43 N - 51 N and 4 W - 6 E, while Scandinavia is defined as a land enclosed between 5 E - 39 E and 57 N - 71 N.  The heatwave duration is chosen depending on the impact of interest. In this study we select $T=\{5,14,30,90\}$ days. In order to remove the climate change signal from ERA5 reanalysis the fields were linearly detrended for each grid point and anomalies were defined with respect to such detrended mean. 

{The choice to study the extremes of temperature anomalies rather absolute temperature is made for the following reason. From a dynamical perspective, the summer months are similar, and provide more statistics than individual months. This definition emphasizes dynamical characteristics, which are thought to be described reasonably well on a seasonal time scale, rather than physical impacts which are often related to the physical temperature. Moreover, the emphasis on anomalies allows to collect richer statistics and to concentrate on properties that are better captured by the models, as opposed to the absolute magnitudes.} 

{We are interested in summer statistics June July August (JJA) so that duration of heatwave is constrained to JJA and cannot occur outside this period.  
The yearly block-maxima extremes $a_{i}$ occurring at calendar day $t_i$ are defined as}
\begin{equation}\label{eq:yearly_extremes}
    a_{i}:=A\left(t_{i}\right)=\max_{t\in\text{summer }(i)}\left\{ A(t) \right\} ,
\end{equation}

Since temperatures are correlated on a daily timescales one has to be careful about the selection of the maxima which occur on the first or the last summer day. Thus, for such events we check whether they correspond to local maxima. If they do not, we look for greatest local maximum in the summer time series for that year, thus avoiding registering extrema of May or September. \footnote{This is because with the original definition any extreme which actually occurs in May or September can superficially inflate the summer extremes for the first day or the last day of the summer because of temporal correlations. }

To compute the return times, the yearly summer extremes $a_{i}$ are ranked in decreasing order, as usual (see Section~\ref{sec:returns}). 
Next, we compute return times for heatwaves using the method~\citet{Lestang_2018} described in supplementary material. The return time expression reads in our notation as
\begin{equation}\label{eq:return_equation}
    r = \frac{1}{\log \left(\text{1 - rank}(a)/M \right)},
\end{equation}
{where $M$ corresponds to the total number of years and $\text{rank}(a)$ corresponds to the index $i$ of the extreme found in the sequence $\left\{ a_{i}\right\} $. 
For very rare values $a$, this expression gives $r \approx M/\text{rank}(a)$, which is the more familiar definition of return time found in the climate literature.}

 \subsection{{Statistical analysis}}\label{sec:statistics}

To identify long lasting heatwaves we will consider time averages of temperature and geopotential height on different time scales. In particular, given a variable $X(\mathbf{x},t)$  
we consider averages conditioned on the extreme, also called composite maps. The composite map of $X$ at the lead time $\tau$ of
an extreme heatwave with a return time $r$ is 
\begin{equation}\label{laggedcomposite}
X^{r}(\mathbf{x},\tau)=\frac{1}{N_r}\sum_{\left\{ i\left|a_{i}\geq a^{r}\right.\right\} }X(\mathbf{x},t_{i}-\tau).
\end{equation}
where $N_r$ is a number of events for a given threshold $a_r$ and {$t_{i}$ is the time of the onset of the heatwave}. This definition gives informations on how a typical heatwave above a given threshold develops and disappears. 


We test statistical significance using the standard Student $t$ test. For instance, let us discuss the estimation of the average of a random variable $Y$, using $N$ independent samples $\{Y_n\}$ through the empirical average 
\begin{equation}
    {\mathbb E}_e(Y) = \frac{1}{N}\sum_{n=1}^N Y_n
\end{equation}
If our aim is to test whether this sample average is significant, in the sense that the probability that the actual average is 0 is very low, we will compute the t-value 
\begin{equation}
    t=\frac{\sqrt{N}{\mathbb E}_e(Y)}{{\sigma_e(Y)}}, 
\end{equation}
where $\sigma_e(Y)$ is the empirical standard deviation
\begin{equation}
    \sigma_e(Y) = \sqrt{ \frac{1}{N-1} \sum_{n=1}^{N}\left[Y_n - {\mathbb E}_e(Y) \right]^{2} }. 
\end{equation}
{We consider the average statistically significant with $|t|\geq 2$. We note here that we have taken a threshold on the t-value rather than on the confidence probability. However our smallest sample size is around 16, and the t-value for a given confidence probability changes very weakly with the sample size above 15-20. In any case with our sample sizes a t-value of 2 is conservatively consistent with 95\% confidence level at the very least ~\citep{storch02}.} 

The method of estimating uncertainties known as \emph{bootstrapping} will be used. This involves breaking the total temporal sequence into 10 subsets and computing the relevant quantities correspondingly. 
Afterwards, mean $\langle x\rangle$ serves as the best estimate, while standard deviation $\sigma_x$ as uncertainty. 



\section{{Hemispheric teleconnection patterns}}



\subsection{Hayashi spectra {of midlatitude atmospheric waves}}

{The extreme teleconnection patterns found by \citet{Ragone18,Ragone:2020vs,Ragone21} in  simulations with climate models of different complexity, suggest that the local atmospheric blockings responsible for heatwaves occurring in subregions of Europe could be associated with stationary or very slowly moving atmospheric waves with wavenumber between 3 and 4. The body of literature of \citet{Petoukhov2013,petoukhov16,petoukhov18,Kornhuber_2019,Kornhuber20} identifies the amplification of quasi-stationary Rossby waves with a different range of wavenumbers, between 5 and 8 as responsible for the occurrence of several observed midlatitude extreme events, including heatwaves.}



Since we are seeking to detect global quasi-stationary Rossby wave patterns we use classical space-time Fourier decomposition into eastward/westward and stationary Rossby waves~\citet{Hayashi71}. This analysis is performed on 500 hPa gepotential height averaged in the meridional direction over a latitudinal band centered over the region affected by the heatwaves. The procedure is done via an assumption about the nature of the waves, such as attributing incoherent part of the spectrum to real travelling waves~\citet{pratt76,Hayashi79}. The low frequency - low wavenumber domain corresponds to standing and westward propagating waves, while the high frequency-high wavenumber to synoptic disturbances.  For the details also see the appendix of~\citet{DellAquila_2005ta}. We chose the latitudinal belt 55–75 N as a region relevant for Scandinavia over which the averaging was performed. 
This belt is part of the larger belt 30-75 N where most of the baroclinic and the low frequency wave activity occurs. Following the procedures in the literature, the meridional integration is an average, not a sum, so the total power does not scale with the size of the latitude band. Additionally, in order to compensate for the non-constant density of points on a log-log plot the spectra are multiplied by $k\cdot \omega/2\pi$.

\setcounter{figure}{0}
\setcounter{subfigure}{0}
\begin{subfigure}
    \centering
    \begin{minipage}[b]{0.48\textwidth}
        \includegraphics[width=\linewidth]{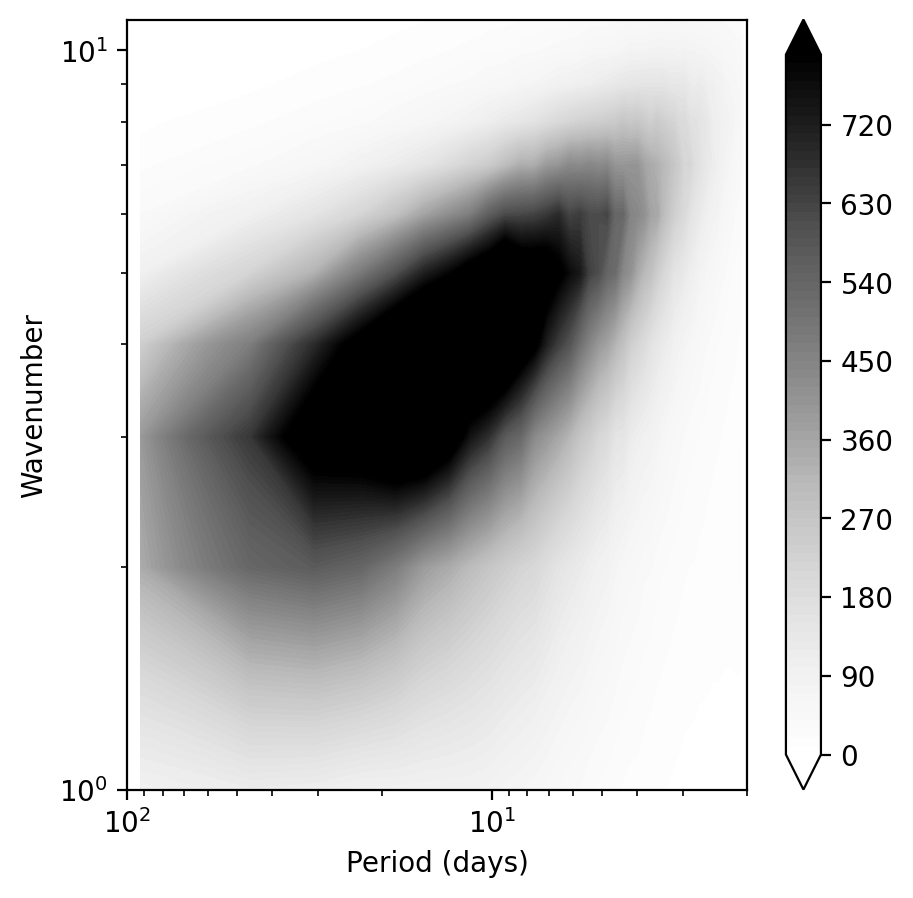}
        \caption{$H_E(k,\omega)$, th = $\infty$.}
        \label{fig:HEinfty}
    \end{minipage}  
    \hfill
\setcounter{figure}{1}
\setcounter{subfigure}{1}
    \begin{minipage}[b]{0.48\textwidth}
        \includegraphics[width=\linewidth]{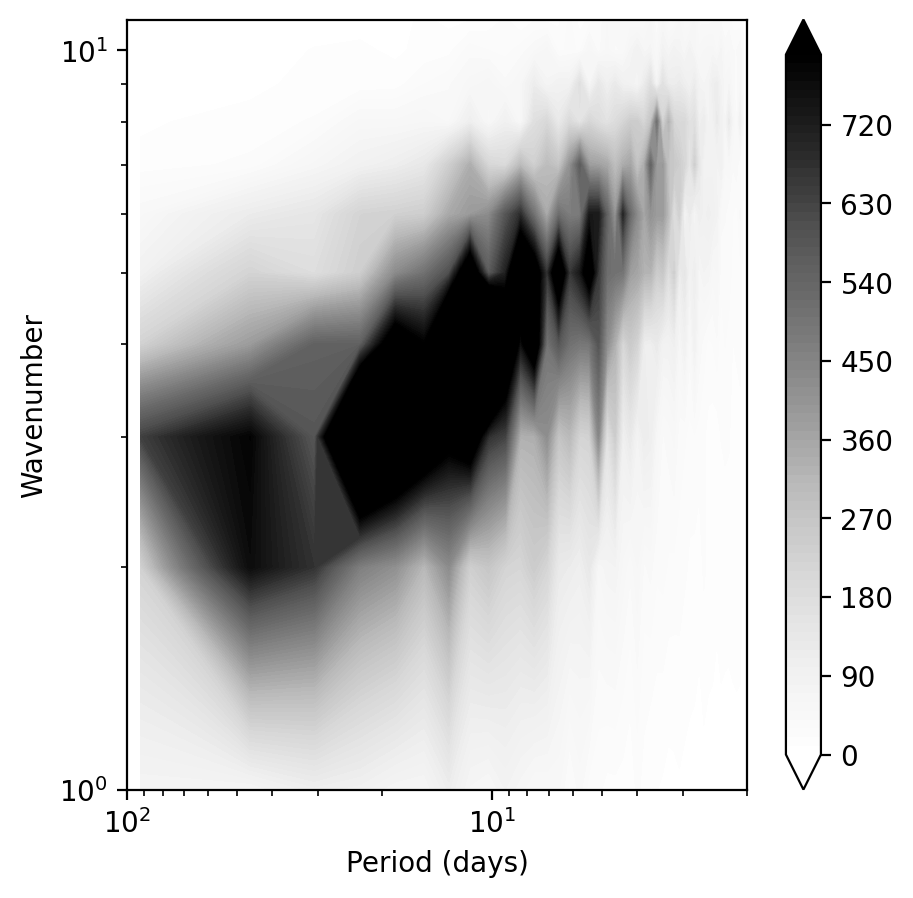}
        \caption{$H_E(k,\omega)$, th = $-4.5$.}
        \label{fig:HEPeak}
    \end{minipage}

\setcounter{figure}{1}
\setcounter{subfigure}{-1}
    \caption{Comparison of Hayashi spectra for \textbf{(A)} 1000 years of CESM climatology and  \textbf{(B)} a subset of 1000 years of CESM with Scandinavian heatwaves above the threshold of 4.5 degrees {of $T = 30$ day heatwaves}. Boundaries of the latitude band over which the meridional average is taken are 55 and 75.}
    \label{fig:HayashiEastward}
\end{subfigure}

In what follows we work with conditional averages (condition is to keep only the years where $A(t)>th$, equation~\eqref{timeaveraged}, for a certain threshold $th$, so that $th = -\infty$ corresponds to climatology). We show Hayashi spectra for different thresholds on Figure~\ref{fig:HayashiEastward}. 
 On the left panel we plot climatology threshold $th = -\infty$, which displays the classical spectrum of Rossby waves. On the right panel we plot Hayashi spectra for the summers with 16 most extreme $T = 30$ day heatwaves, which corresponds to the threshold $th= 4.5 K$. We observe a structure at a period of order 50 days and wavenumber 3 for eastward propagating waves. We note that this pattern in the Hayashi spectrum is isolated from the rest of the Rossby wave branch and it suggests that quasi-stationary processes take place when conditioned to long-lasting heatwaves. 

\subsection{Extreme teleconnection pattern}\label{scandinavian2018}

\setcounter{figure}{2}
\setcounter{subfigure}{0}
\begin{subfigure}
    \centering
    \begin{minipage}[b]{0.3\textwidth}
        \includegraphics[width=\linewidth]{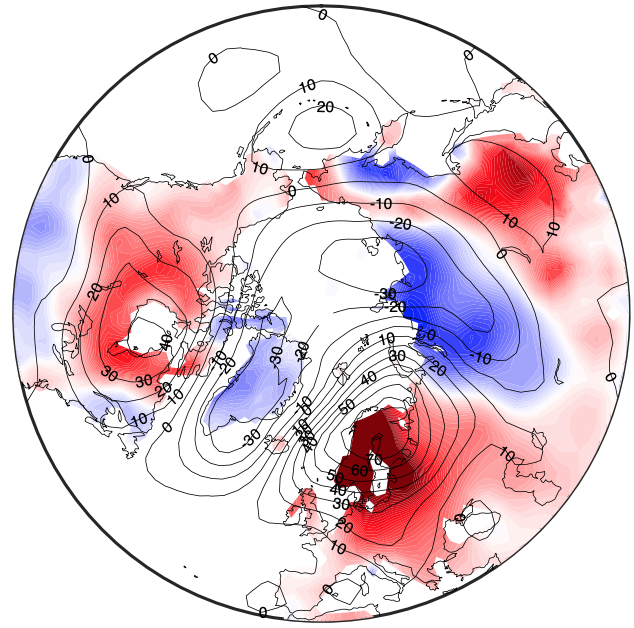}
        \caption{PlaSim rare event}
        \label{fig:PNASPlaSimTripole}
    \end{minipage}  
    \hfill
\setcounter{figure}{2}
\setcounter{subfigure}{1}
    \begin{minipage}[b]{0.3\textwidth}
        \includegraphics[width=\linewidth]{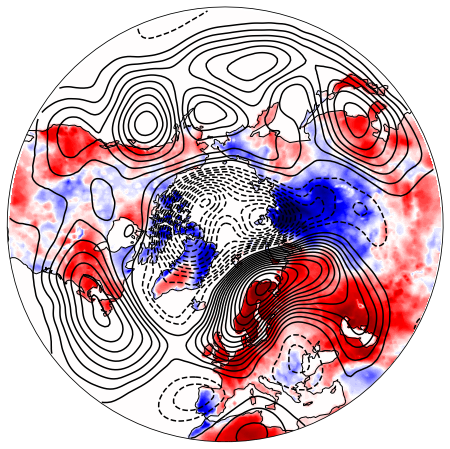}
        \caption{ERA5 reanalysis}
        \label{fig:ERA5ScandinaviaTripole}
    \end{minipage}
    \hfill
\setcounter{figure}{2}
\setcounter{subfigure}{2}
    \begin{minipage}[b]{0.35\textwidth}
        \includegraphics[width=\linewidth]{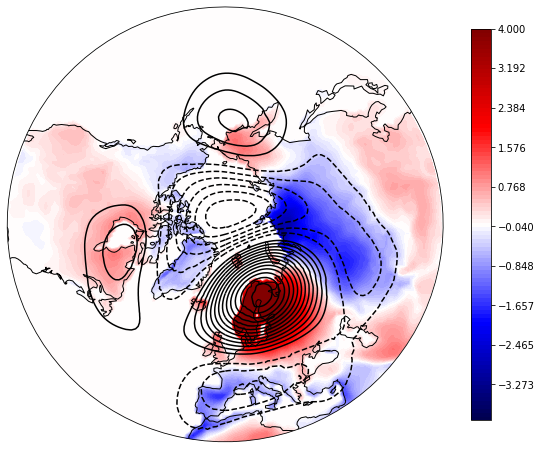}
        \caption{CESM 1000 years}
        \label{fig:CESMScandinaviaTripole}
    \end{minipage}

\setcounter{figure}{2}
\setcounter{subfigure}{-1}
    \caption{\textbf{(A)} Northern Hemisphere surface temperature anomaly (colors) and 500-hPa geopotential height anomaly (contours), conditional on the occurrence of European heatwaves with $T = 90$ days and the threshold of $a = 2 K$, estimated from the large deviation algorithm, courtesy to \citet{Ragone18}. \textbf{(B)}  ERA5 reanalysis averaged over July 2018 and detrended relative to the 1950-2020 linear trend, we follow the same conventions for representing the geopotential height and temperature anomalies, except that negative geopotential anomalies are presented via dashed lines. \textbf{(C)} Composite map of CESM drawn from 1000 year long run and conditioned to $T = 30$ day heatwaves above the threshold of 3.5 degrees. We follow the same conventions for representing the geopotential height and temperature anomalies. The temperature colormap is shared among the three plots.  }
    \label{fig:Scandinavia2018}
\end{subfigure}

Hayashi spectra associated with heatwaves in Scandinavia, computed with the CESM model, show a clear local, low-frequency, low wave-number peak (See Figure~\ref{fig:HayashiEastward}). There are two questions that arise immediately:
\begin{itemize}
    \item What is the real-space pattern that corresponds to this peak?
    \item Does the teleconnection pattern have a counterpart in the the observational record
\end{itemize}
The chief quantities of interest are 500 hPa geopotential height (500GPH) and the 2 meter temperature (T2M).The 500GPH is a convenient quantity to analyse the large scale circulation in the troposphere where the effects of the topography are minor \citet{Blackmon76}.  

To answer the first question we compute composite maps (see equation~\eqref{laggedcomposite}) of 500GPH and T2M on Figure~\ref{fig:Scandinavia2018} conditioned to the $T = 30$ day long heatwaves of Scandinavia of threshold $a = 3.5 K$ (containing 65 events). {The choice of the threshold is motivated by the comparisons with the actual event which was observed in reanalysis that will be commented on below}. We observe a tripole-like structure, which consists of an Arctic cyclonic anomaly surrounded by three anti-cyclonic anomalies, over Scandinavia, Chukotka and North-East Canada. The cyclonic anomaly extends to Greenland and Siberia, and it is associated with negative temperature anomalies. This teleconnection pattern indeed corresponds to the local maximum in the low-frequency low-wavenumber branch of the Hayashi spectrum (Figure~\ref{fig:HEPeak}). 

Motivated by the second question we select the July 2018 event from ERA5 reanalysis. The details concerning the data processing of ERA5 are discussed in the Section~\ref{era5reanalysis}. The July 2018 event consists of a prolonged heatwave in Scandinavia, whose temperature anomaly was above $3 - 3.5$ degrees for about a month. Its composite map (Figure~\ref{fig:ERA5ScandinaviaTripole})  displays various similarities with the CESM teleconnection (Figure~\ref{fig:CESMScandinaviaTripole}). Most striking resemblance are the negative Arctic 500GPH anomaly, as well as positive anomalies in Scandinavia and North-East Canada. Reanalysis favors a positive 500GPHA shifted eastwards towards the Atlantic. In the Pacific ocean we see a train of 500GPHA  instead of a single positive anomaly. In general, one finds that geopotential anomalies are stronger in reanalysis; for instance, 500GPHA over Scandinavia is about $140 m$, whereas in CESM we find values on the order of $70 m$.  {The stronger  anomalies can be attributed to the fact that in reanalysis we were averaging over the single event, while in CESM the averaging is performed over multiple heatwaves.} 

We are now in a position to also comment on the relative similarity between these two teleconnection patterns with the one presented in~\citet{Ragone18}. For  illustration purposes we display the copy of the relevant teleconnection on Figure~\ref{fig:PNASPlaSimTripole}. It was obtained using the rare event algorithm for PlaSim conditioned to $T = 90$ day heatwaves of threshold $a = 2 K$ occurring in Europe  (to be more precise each grid point is averaged over 90 days). It turns out that the main contribution is due to heatwaves in Scandinavia, of threshold that can be estimated as $a \sim 3.5 - 4 K$. It is interesting that the similarities between the panels exist despite the different complexity of the two models, PlaSim and CESM. This gives us indication that such teleconnection patterns could be robust feature of the large scale dynamics. However, the specific conditions are not identical, and the similarity to the July 2018 event could be accidental. {For instance, the high temperature and anti-cyclonic anomaly in the southeastern Asia is consistently reproduced in PlaSim and ERA5 but not in CESM}. In order to better test {the assertion that planetary-scale teleconnections are consistent}, we perform a more extensive intra-model comparison in the next Section.


\subsection{Teleconnections in models versus reanalysis}

\setcounter{figure}{3}
\setcounter{subfigure}{0}
\begin{subfigure}
    \centering
    \begin{minipage}[b]{0.32\textwidth}
        \includegraphics[width=\linewidth]{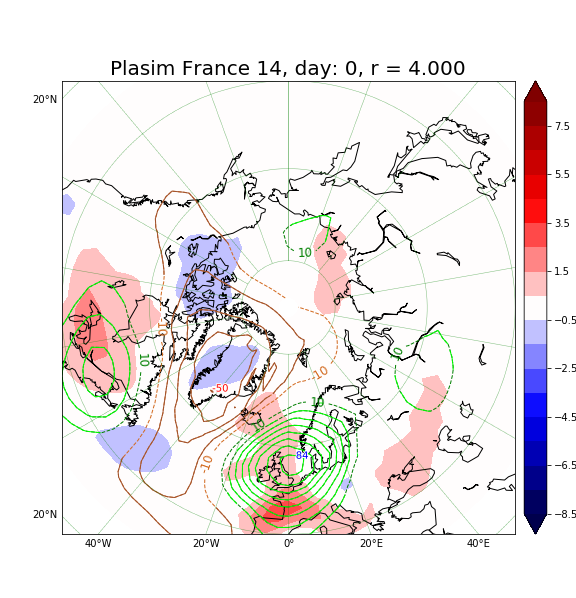}
        \caption{PlaSim France}
        \label{fig:PlasimFrance}
    \end{minipage}  
    \hfill
\setcounter{figure}{3}
\setcounter{subfigure}{1}
    \begin{minipage}[b]{0.32\textwidth}
        \includegraphics[width=\linewidth]{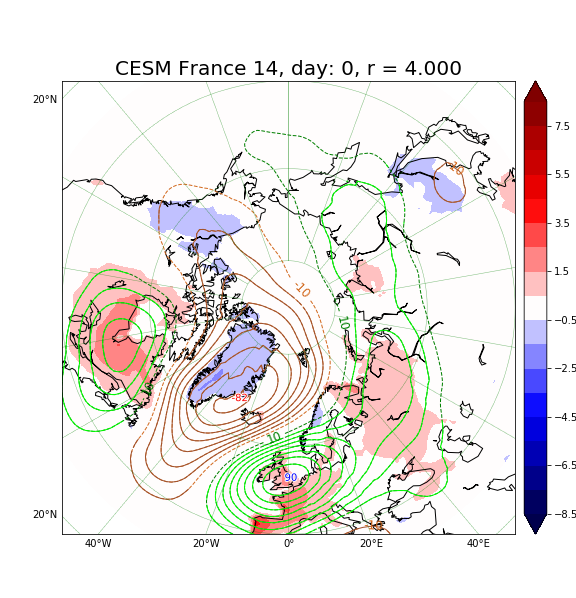}
        \caption{CESM France}
        \label{fig:CESMFrance}
    \end{minipage}
    \hfill
\setcounter{figure}{3}
\setcounter{subfigure}{2}
    \begin{minipage}[b]{0.32\textwidth}
        \includegraphics[width=\linewidth]{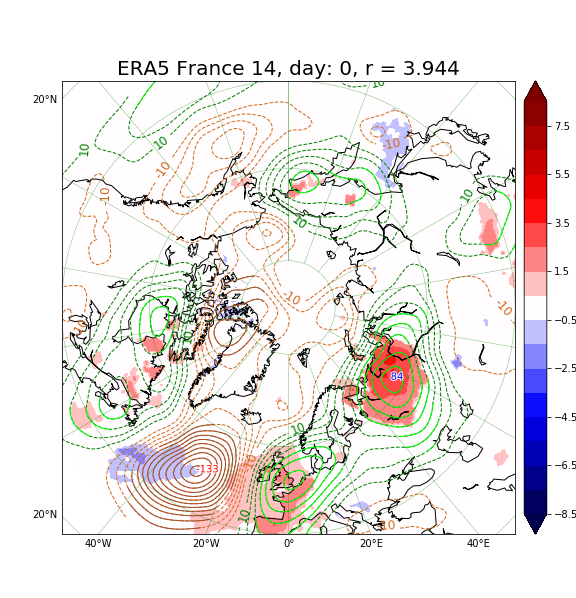}
        \caption{ERA5 France}
        \label{fig:ERA5France}
    \end{minipage}

\setcounter{figure}{3}
\setcounter{subfigure}{3}
    \begin{minipage}[b]{0.32\textwidth}
        \includegraphics[width=\linewidth]{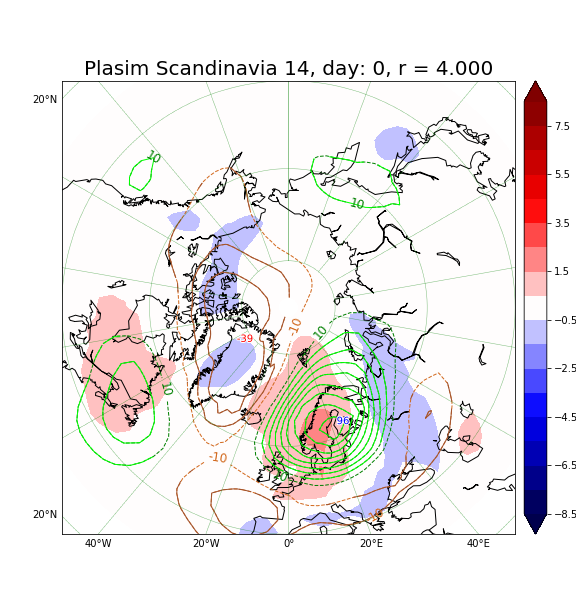}
        \caption{PlaSim Scandinavia}
        \label{fig:PlasimScandinavia}
    \end{minipage}  
    \hfill
\setcounter{figure}{3}
\setcounter{subfigure}{4}
    \begin{minipage}[b]{0.32\textwidth}
        \includegraphics[width=\linewidth]{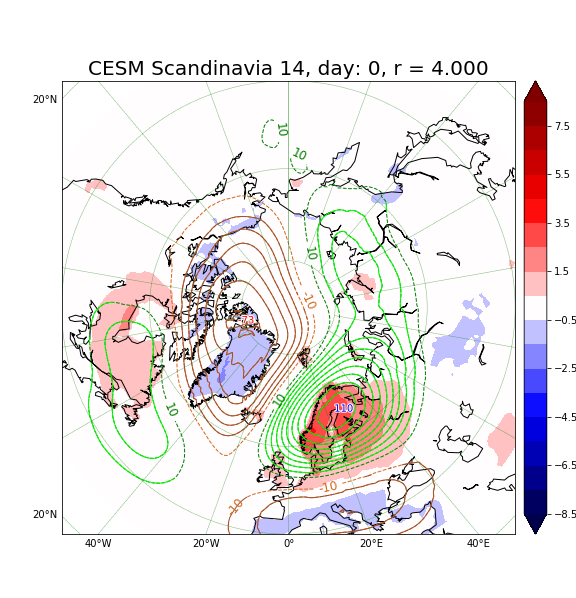}
        \caption{CESM Scandinavia}
        \label{fig:CESMScandinavia}
    \end{minipage}
    \hfill
\setcounter{figure}{3}
\setcounter{subfigure}{5}
    \begin{minipage}[b]{0.32\textwidth}
        \includegraphics[width=\linewidth]{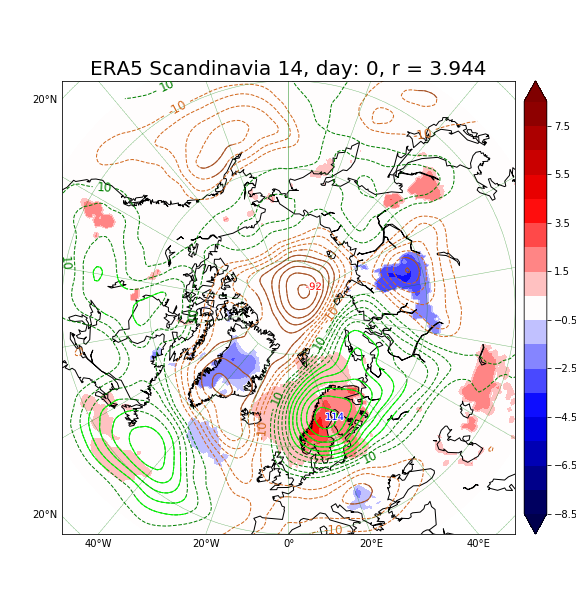}
        \caption{ERA5 Scandinavia}
        \label{fig:ERA5Scandinavia}
    \end{minipage}

\setcounter{figure}{3}
\setcounter{subfigure}{-1}
    \caption{Composites of $r = 4$ year return time in different models (left to right), where 2m temperature and 500 mbar geopotential anomalies are plotted conditioned to different areas (top-bottom):  
    \textbf{(A)} PlaSim France. 
    \textbf{(B)} CESM France. 
    \textbf{(C)} ERA5 France. 
    \textbf{(D)} PlaSim Scandinavia. 
    \textbf{(E)} CESM Scandinavia. 
    \textbf{(F)} ERA5 Scandinavia. at $\tau = 0$ lag (onset of the heatwave). The colormap for the temperature is indicated next to the figures. Only $|t|>2$ statistically significant anomalies of temperature are displayed (see the colormap in K), whereas all level sets of geopotential height anomalies graded by $10 m$ are shown (Green - positive, Orange - Negative). The values of geopotential that are statistically significant $|t| > 2$  are plotted with solid lines and otherwise with dashed lines (see Section~\ref{sec:statistics} on statistical significance). The global maximum and minimum of geopotential is shown using blue and red text tooltips respectively. All subsequent figures displaying composites will follow this general style.
    }
    \label{fig:comp_tau0}
\end{subfigure}

We aim to provide systematic comparisons between the models. We concentrate on two areas in Europe, corresponding to France and Scandinavia. The areas have been chosen to correspond closely to the heatwave clusters in Western Europe and Scandinavia identified by~\citet{Stefanon_2012}.

Since the ERA5 dataset contains only 71 years we limit the study to return times not larger than 4 years, in order to have enough data (at least 18 heatwaves). For consistency we select heatwaves with the same return time of 4 years in CESM and PlaSim (250 heatwave events in both models) which allows for stronger statistical inference. The corresponding thresholds $a$ are not the same across the models, especially in Scandinavia, where the threshold for PlaSim is much lower than the other two datasets (see table~\ref{tab:4yearthresholds}). This is further discussed in Section~\ref{sec:returns}.
\begin{table}
    \begin{center}
     \begin{tabular}{||l c c ||} 
     \hline
      & France & Scandinavia  \\ [0.5ex] 
     \hline\hline
     PlaSim & $3.23K$ & $1.74 K$  \\ 
     \hline
     CESM & $3.47K$ & $3.79 K$ \\
     \hline
     ERA5 & $3.45K$ & $3.25 K$ \\[1ex]
    \end{tabular}
    \end{center}
    \caption{Values of a $r=4$ year return time threshold $a_4$ computed for France and Scandinavian heatwaves (columns) over PlaSim, CESM and ERA5 datasets (raws). }
    \label{tab:4yearthresholds}
\end{table}

\setcounter{figure}{4}
\setcounter{subfigure}{0}
\begin{subfigure}
    \centering
    \begin{minipage}[b]{0.48\textwidth}
        \includegraphics[width=\linewidth]{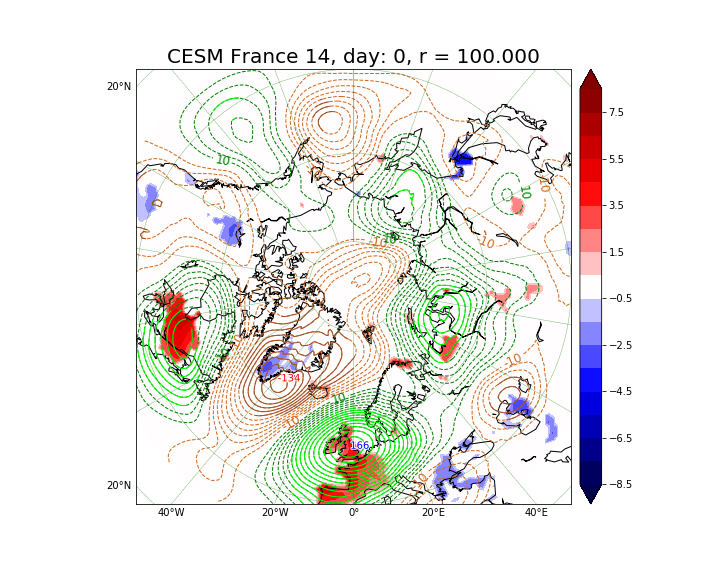}
        \caption{$r>100$ years}
        \label{fig:r100}
    \end{minipage}  
    \hfill
\setcounter{figure}{4}
\setcounter{subfigure}{1}
    \begin{minipage}[b]{0.48\textwidth}
        \includegraphics[width=\linewidth]{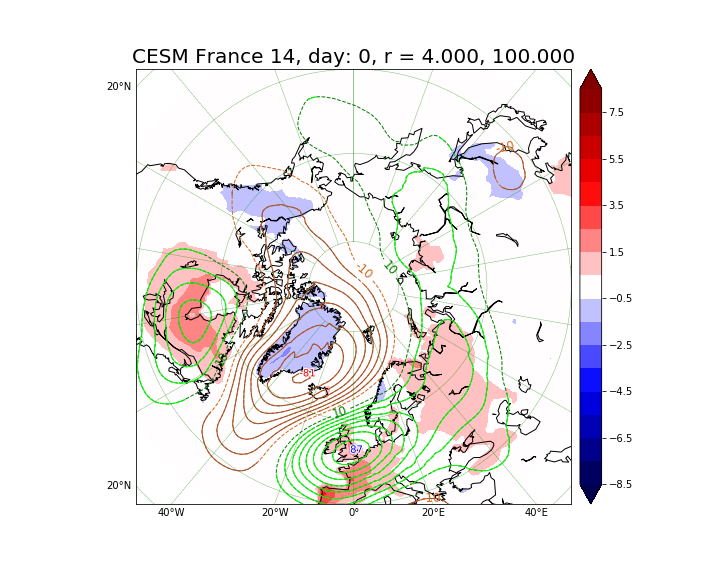}
        \caption{$4<r<100$ years}
        \label{fig:4r100}
    \end{minipage}

\setcounter{figure}{4}
\setcounter{subfigure}{-1}
    \caption{Comparison between \textbf{(A)} $r > 100$  and \textbf{(B)} $4 < r < 100$ for France heatwaves in CESM dataset. See the caption of Figure~\ref{fig:comp_tau0} for plotting conventions.}
    \label{fig:comp_r100vsr4to100}
\end{subfigure}

One wonders how similar are the teleconnection patterns drawn from models at the onset of individual heatwave events, i.e. at $\tau = 0$ days. We take duration $T = 14$ days, and plot the corresponding composites on Figure~\ref{fig:comp_tau0}, following the method described in Section~\ref{methodology}, for heatwaves occurring in France (top) and Scandinavia (bottom). We clearly see the presence of a tripole-like structure in all the images, but there are many additional details. PlaSim and CESM share more features, although qualitative similarities can also been seen when comparing with ERA5. In particular, one observes statistically significant positive 500GPHA in the North sea of order 84 m in PlaSim, 90 m in CESM and 50 m in ERA5. The second largest in magnitude feature is the negative Arctic 500GPHA of order -50 m in PlaSim and -82 m in CESM which extends to the Atlantic. The extension of this pattern seems to be consistent with what one observes in ERA5. However, in ERA5 it is not statistically significant. In contrast, we have a statistically significant negative 500GPHA in the Atlantic ocean of -133 m, which is not so strong in PlaSim or CESM. The third order feature is the positive 500GPHA in East Canada which reaches 30 m in CESM and PlaSim and approximately 40 m in ERA5. Finally, in PlaSim we have 10 m in North-East Asia anomalies, which are stronger in CESM and ERA5. Most features are statistically significant for CESM and PlaSim but many are not for ERA5 (see Section~\ref{sec:statistics} on statistical significance test), which is due to scarcity of the observational record. 

The panels at the bottom of Figure~\ref{fig:comp_tau0} corresponding to teleconneciton patterns conditioned to Scandinavian heatwaves also show of these three features: a \emph{primary} one consisting of positive 500GPHA in Scandinavia: 96 m in PlaSim, 110 m in Scandinavia and 114 m in ERA5; a \emph{secondary} one consisting of negative 500GPHA in the Arctic with -39 m in PlaSim, -73 m in CESM, and in ERA5 of order - 90m. While this peak is larger than expected, we find that overall shape and magnitude of 500GPHA in the Arctic are qualitatively consistent between the models and reanalysis. 
The \emph{tertiary} pattern consists of positive 500GPHA in Eastern Canada which is approximately 20 m in both PlaSim and CESM and of order 60 m in ERA5 with a easterly shift.  Finally there is also a pattern which seems to be generally consistent across the models and reanalysis: the Mediterranean depression of order -20 m. However, in reanalysis it is, again, not statistically significant. 

To summarize, we find qualitative agreement between Figures~\ref{fig:PlasimFrance},~\ref{fig:CESMFrance} and~\ref{fig:ERA5France} as well as agreement between~\ref{fig:PlasimScandinavia},~\ref{fig:CESMScandinavia} and~\ref{fig:ERA5Scandinavia}. CESM seems to be capturing the teleconnection patterns closer to reanalysis. For instance, in case of heatwaves conditioned to France we see that CESM captures better the distribution of geopotential anomalies over Canada and Northern Russia when compared to ERA5. Nevertheless, the similarities displayed in the teleconnection patterns suggest that they are model independent and have observational counterparts that are rather robust.

\setcounter{figure}{5}
\setcounter{subfigure}{0}
\begin{subfigure}
    \centering
    \begin{minipage}[b]{0.48\textwidth}
        \includegraphics[width=\linewidth]{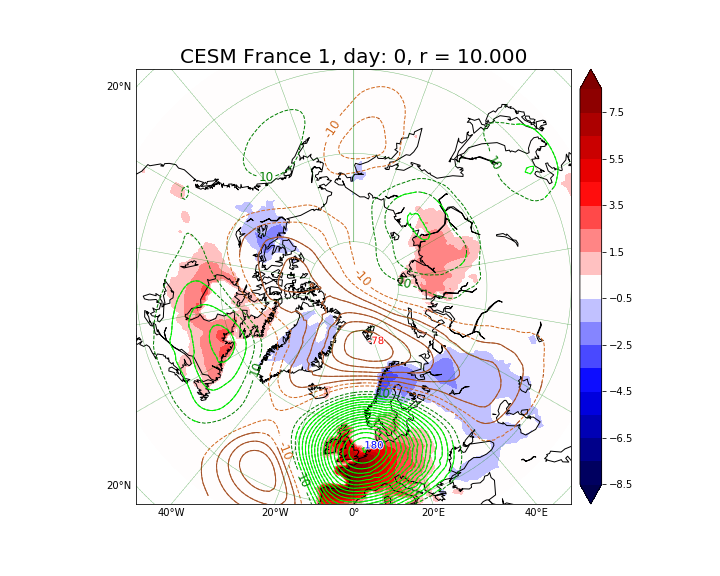}
        \caption{$T = 1$ days}
        \label{fig:T1}
    \end{minipage}  
    \hfill
\setcounter{figure}{5}
\setcounter{subfigure}{1}
    \begin{minipage}[b]{0.48\textwidth}
        \includegraphics[width=\linewidth]{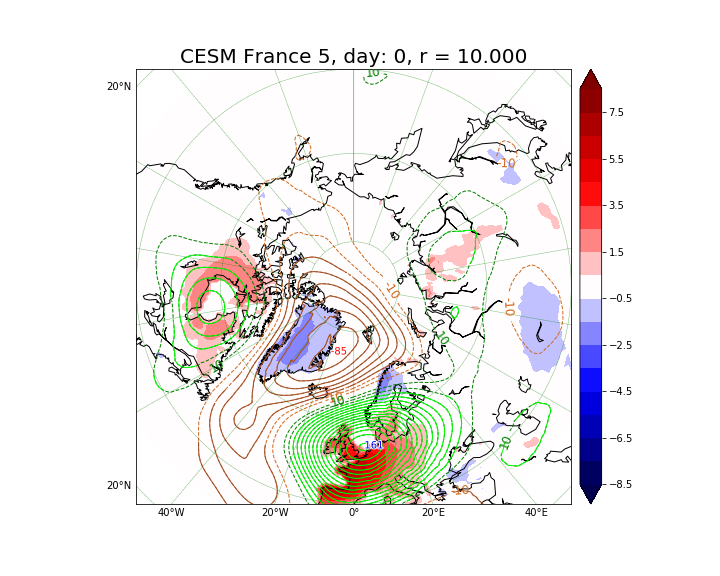}
        \caption{$T = 5$ days}
        \label{fig:T5}
    \end{minipage}

\setcounter{figure}{5}
\setcounter{subfigure}{2}
    \begin{minipage}[b]{0.48\textwidth}
        \includegraphics[width=\linewidth]{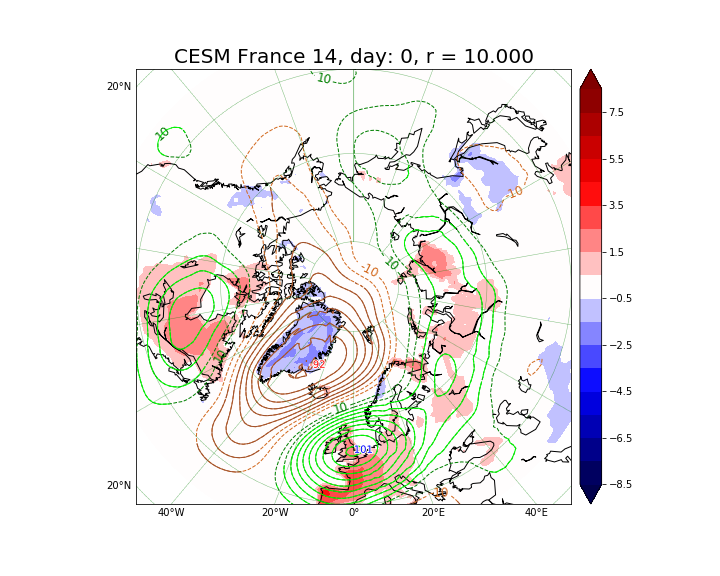}
        \caption{$T = 14$ days}
        \label{fig:T14}
    \end{minipage}  
    \hfill
\setcounter{figure}{5}
\setcounter{subfigure}{3}
    \begin{minipage}[b]{0.48\textwidth}
        \includegraphics[width=\linewidth]{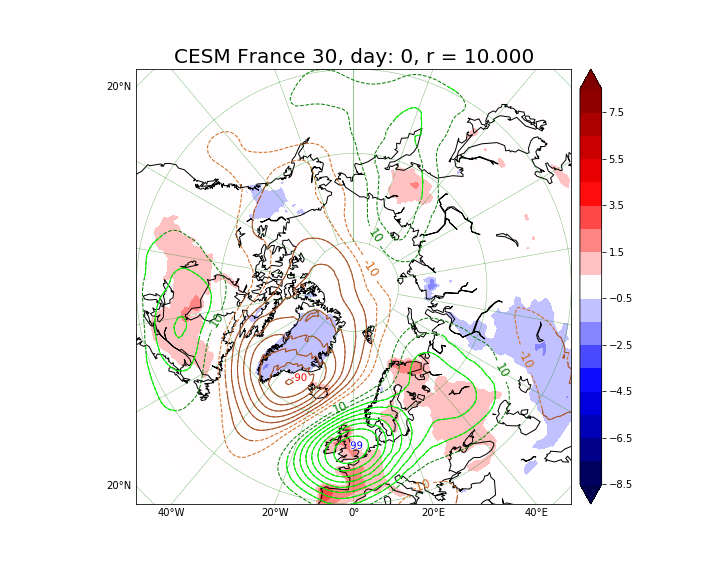}
        \caption{$T = 30$ days}
        \label{fig:T30}
    \end{minipage}

\setcounter{figure}{5}
\setcounter{subfigure}{-1}
    \caption{Composites of 2m temperature and 500 mbar geopotential conditioned on $r\ge 10$ year return time heatwaves over France at $\tau = 0$ lag. The reference model is CESM. Among the different panels we vary the length of heatwaves: \textbf{(A)} $T = 1$, \textbf{(B)} $T = 5$, \textbf{(C)} $T = 14$ and \textbf{(D)} $T = 30$ days. See the caption of Figure~\ref{fig:comp_tau0} for further plotting conventions. }
    \label{fig:comp_tau0_times}
\end{subfigure}

Another important question is how these patterns change when conditioning for more extreme heatwaves. In Figure~\ref{fig:comp_r100vsr4to100} we compare extreme heatwaves with return time larger than 100 years, with the ones with return time in the range $4 < r < 100$. In the latter case we have actually removed the 10 most extreme events to show that they do not contribute noticeably compared to the remaining 240 events. Like in the previous analysis we identify the three most important features as {negative GPHA over Greenland} and positive GPHA over North sea and East Canada. In Figure~\ref{fig:r100}, for the events that are more extreme, they take values 166 m, -134 m and approximately 90 m, while on Figure~\ref{fig:4r100}, for the events that are milder, they take the values 87 m, -81m and approximately 40 m. Furthermore, Figure~\ref{fig:r100} has variety of other mostly statistically insignificant cyclonic/anticyclonic anomalies.  The conclusion is that when going from less extreme to more extreme heatwaves, the most significant features remain geographically fixed (500GPHA naturally deepen), while we obtain a few more anomalies that do not pass $|t| > 2$ test. {For comparisons with Scandinavia see Figure S1 in supplementary material.}

Finally, we assess how the teleconnection patterns depend on the choice of the heatwave duration $T$ (see equation~\eqref{timeaveraged}). We test 4 periods, $T = \{1, 5,14,30 \}$ days for $r = 10$ year return times, that are shown in Fig.~\ref{fig:comp_tau0_times} for the CESM dataset conditioned to heatwaves in France. 
The patterns discussed before, including the nodes of the tripole structure, such as positive GPHA on the North sea, Canada and East Siberia can be recovered also from Figure~\ref{fig:T5} and Figure~\ref{fig:T30} with mostly the same intensities and some minor shifts. Most different of all is Figure~\ref{fig:T1} which is actually a daily average. In this case the Arctic anomaly descends towards Scandinavia/western Russia region and we have an isolated negative GPHA in the Atlantic. The differences are to be expected since on daily time scale we may observe many of the synoptic features. We conclude that the features we have described in CESM at $\tau = 0$ appear to be also robust across different time-averaging intervals, namely at $T = \{5,14,30 \}$ days. However, we observe that for shorter periods such as 1-5 days negative anomaly in the Arctic develops tongues that reach the midlatitudes. These are not present for large values of $T$ as the faster synoptic type perturbations are averaged over.  {For comparisons with Scandinavia see Figure S2 in supplementary material.}

\section{Temporal evolution of the extremes}

The goal of this Section is to compare the temporal statistics between the reanalysis and the models. The time evolution of composite statistics will only be considered for CESM, whereas other types of statistical properties will be compared across the models. 

\subsection{The dynamical evolution of composites}

Here we address the dynamical evolution leading up to and during the heatwave. We limit this study to the CESM model because we lack data in ERA5 reanalysis and CESM is a higher fidelity model than PlaSim. In Figure~\ref{fig:comp_tau4_movie} we show the dynamical evolution of heatwave composites, conditioned to $r\ge 4$, in France, and varying the lag time $\tau$. This shows how the teleconnection patterns change as a function of lead time. We see that 15 days prior to the onset of heatwaves (Figure~\ref{fig:tau_15}) we already have significant negative GPHA of order -27 m in the Arctic region that slowly broadens and deepens ($-35 m$ at $\tau=-6$ days, Figure~\ref{fig:tau_6}, and $-67~m$ at $\tau = -3$ days, Figure~\ref{fig:tau_3}). Meanwhile, a 
anticyclonic block develops at the midlatitudes, reminiscent of a similar pattern in Figure~\ref{fig:T14}. Five days after the onset (Figure~\ref{fig:tau5}) we can distinguish the tripole-structure (see Section~\ref{scandinavian2018}) with strong significant 133 m positive GPHA in the North sea and negative GPHA that attains -76 m in the North Pole. This pattern starts to disappear at $\tau=15$ days (Figure~\ref{fig:tau15}), corresponding to the duration of the heatwaves.

\setcounter{figure}{6}
\setcounter{subfigure}{0}
\begin{subfigure}
    \centering
    \begin{minipage}[b]{0.32\textwidth}
        \includegraphics[width=\linewidth]{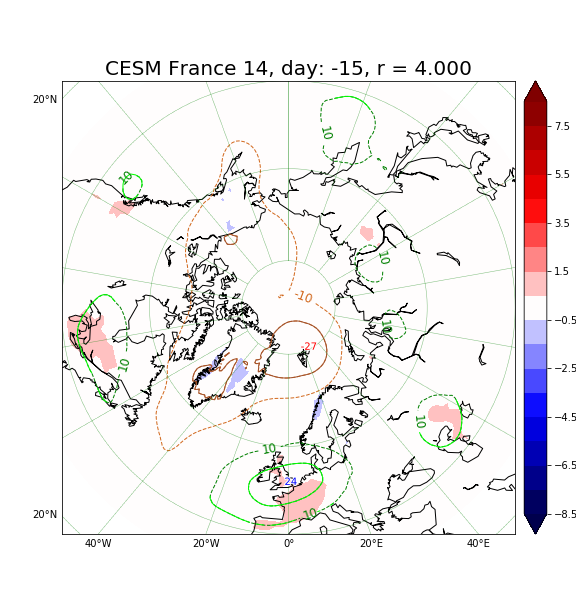}
        \caption{$\tau = -15$ days}
        \label{fig:tau_15}
    \end{minipage}  
    \hfill
\setcounter{figure}{6}
\setcounter{subfigure}{1}
    \begin{minipage}[b]{0.32\textwidth}
        \includegraphics[width=\linewidth]{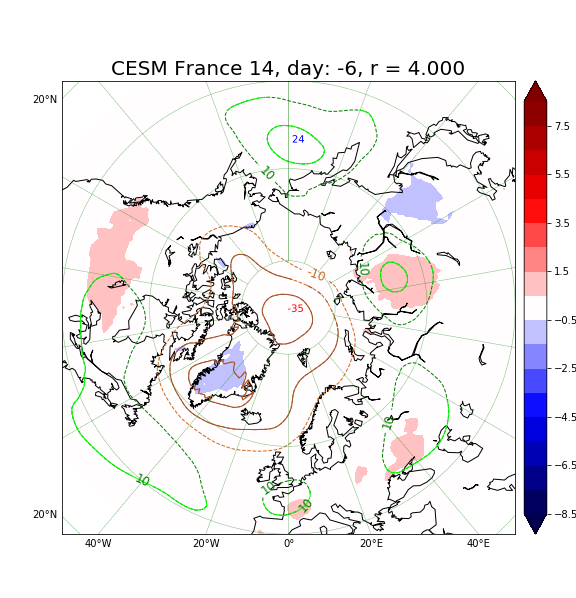}
        \caption{$\tau = -6$ days}
        \label{fig:tau_6}
    \end{minipage}
    \hfill
\setcounter{figure}{6}
\setcounter{subfigure}{2}
    \begin{minipage}[b]{0.32\textwidth}
        \includegraphics[width=\linewidth]{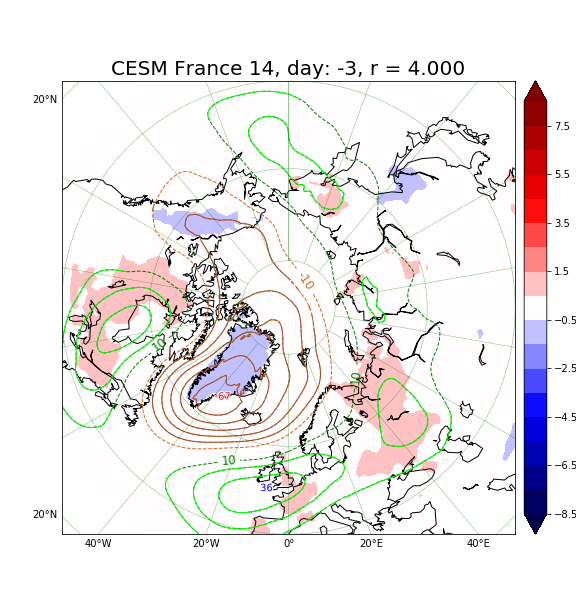}
        \caption{$\tau = -3$ days}
        \label{fig:tau_3}
    \end{minipage}

\setcounter{figure}{6}
\setcounter{subfigure}{3}
    \begin{minipage}[b]{0.32\textwidth}
        \includegraphics[width=\linewidth]{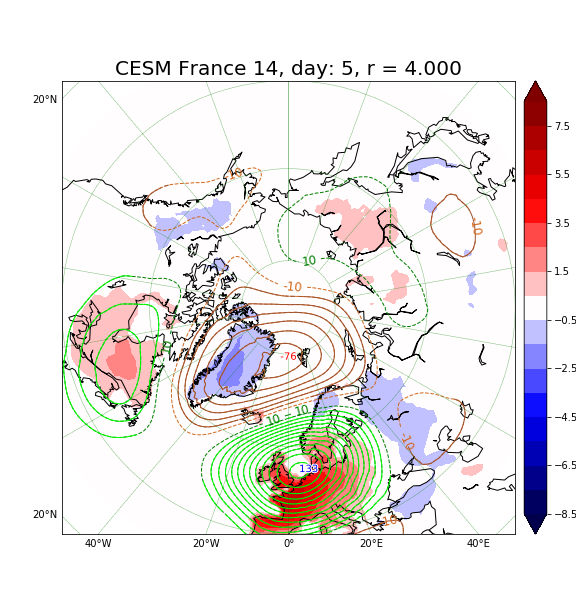}
        \caption{$\tau = 5$ days}
        \label{fig:tau5}
    \end{minipage}  
    \hfill
\setcounter{figure}{6}
\setcounter{subfigure}{4}
    \begin{minipage}[b]{0.32\textwidth}
        \includegraphics[width=\linewidth]{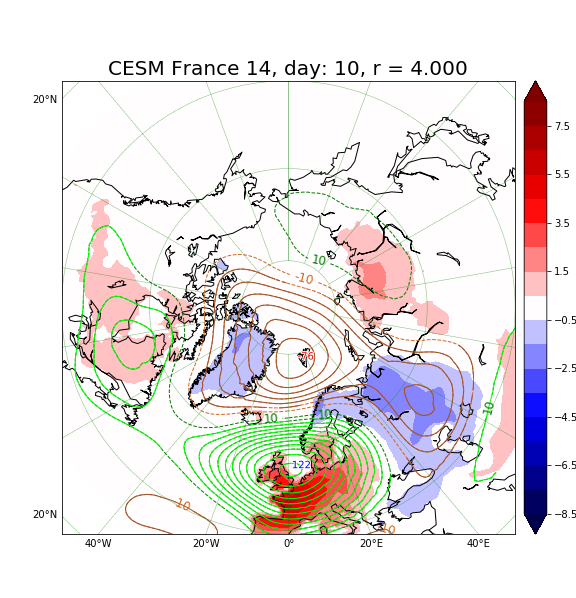}
        \caption{$\tau = 10$ days}
        \label{fig:tau10}
    \end{minipage}
    \hfill
\setcounter{figure}{6}
\setcounter{subfigure}{5}
    \begin{minipage}[b]{0.32\textwidth}
        \includegraphics[width=\linewidth]{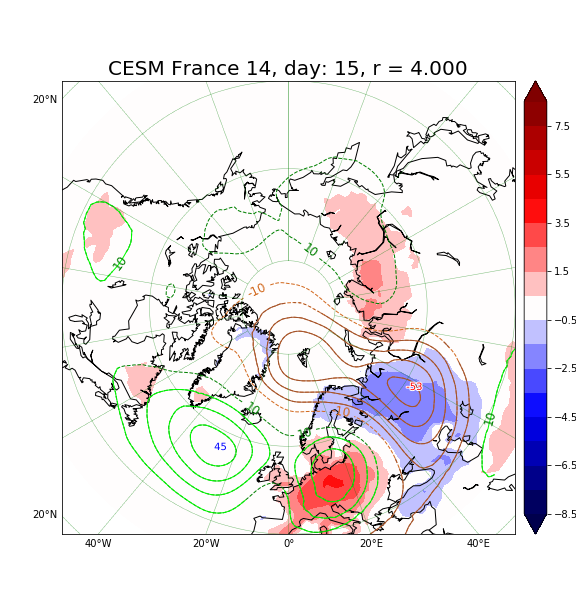}
        \caption{$\tau = 15$ days}
        \label{fig:tau15}
    \end{minipage}

\setcounter{figure}{6}
\setcounter{subfigure}{-1}

    \caption{ Composites of 2m temperature and 500 mbar geopotential conditioned on $r\ge 4$ year return time heatwaves in France at $\tau = 0$ lag. The reference model is CESM and $T = 14$ days is chosen. Control parameter for this figure is $\tau$ with \textbf{(A)} $\tau = -15$, \textbf{(B)} $\tau = -6$, \textbf{(C)} $\tau = -3$,  \textbf{(D)} $\tau = 5$, \textbf{(E)} $\tau = 10$ and \textbf{(F)} $\tau = 15$ days, thus time evolution of CESM composite heatwave in France is displayed. See the caption of Figure~\ref{fig:comp_tau0} for plotting conventions. }
    \label{fig:comp_tau4_movie}

\end{subfigure}

\subsection{Time series analysis}


We discuss the properties of the time series of the area averaged temperature, over France and Scandinavia separately. The idea is to show how closely the models can resemble the real time series and, further, whether we can reduce the dynamics to a simple stochastic process.

\subsubsection{Return time plots}\label{sec:returns}
\setcounter{figure}{7}
\setcounter{subfigure}{0}
\begin{subfigure}
    \centering
    \begin{minipage}[b]{0.45\textwidth}
        \includegraphics[width=\linewidth]{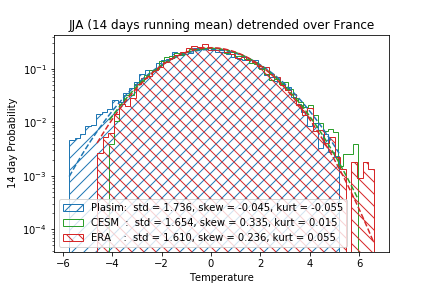}
        \caption{France}
        \label{fig:FranceDist}
    \end{minipage}  
    \hfill
\setcounter{figure}{7}
\setcounter{subfigure}{1}
    \begin{minipage}[b]{0.45\textwidth}
        \includegraphics[width=\linewidth]{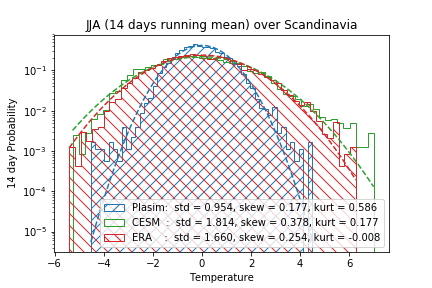}
        \caption{Scandinavia}
        \label{fig:ScandinaviaDist}
    \end{minipage}

\setcounter{figure}{7}
\setcounter{subfigure}{-1}

    \caption{
14 day running mean distributions of daily two meter temperature anomaly (T2MA) integrated over the area of \textbf{(A)} France, \textbf{(B)} Scandinavia, in 3 different datasets: (blue) JJA 1000 year dataset of PlaSim, (green) JJA 1000 year dataset of CESM, (red) JJA 1950-2020 ERA5 reanalysis dataset. We see that France data follows roughly Gaussian statistics, while in Scandinavia PlaSim and to some extent CESM favor non-Gaussianity. The values of standard deviation (std), skewness (skew) and kurtosis (kurt) are given in the inset for each color-coded distribution.
     }
    \label{fig:distributions}

\end{subfigure}

We show the probability distributions of the 14 day running mean of two meter temperature anomaly (T2MA) over France and Scandinavia on Figure~\ref{fig:distributions}. Both areas have Probability Distribution Functions (PDFs) that are Gaussian in the bulk, with fatter tails, particularly for PlaSim (see Section~\ref{gauss_procc} for more details on this). In general, reanalysis datasets are characterized by a noticeable skewness but relatively small kurtosis, which is captured well by CESM in both areas. PlaSim, on the other hand, does not represent well PDF of ERA5 in Scandinavia. Nevertheless, Gaussian statistics seems a reasonable initial assumption as we shall see in Section~\ref{gauss_procc}. This is consistent with the fact that we perform long-time large-area averages: time-averaged statistics that are long enough should assymptote to Gaussian distribution, assuming they are not too correlated. 


\setcounter{figure}{8}
\setcounter{subfigure}{0}
\begin{subfigure}
    \centering
    \begin{minipage}[b]{0.48\textwidth}
        \includegraphics[width=\linewidth]{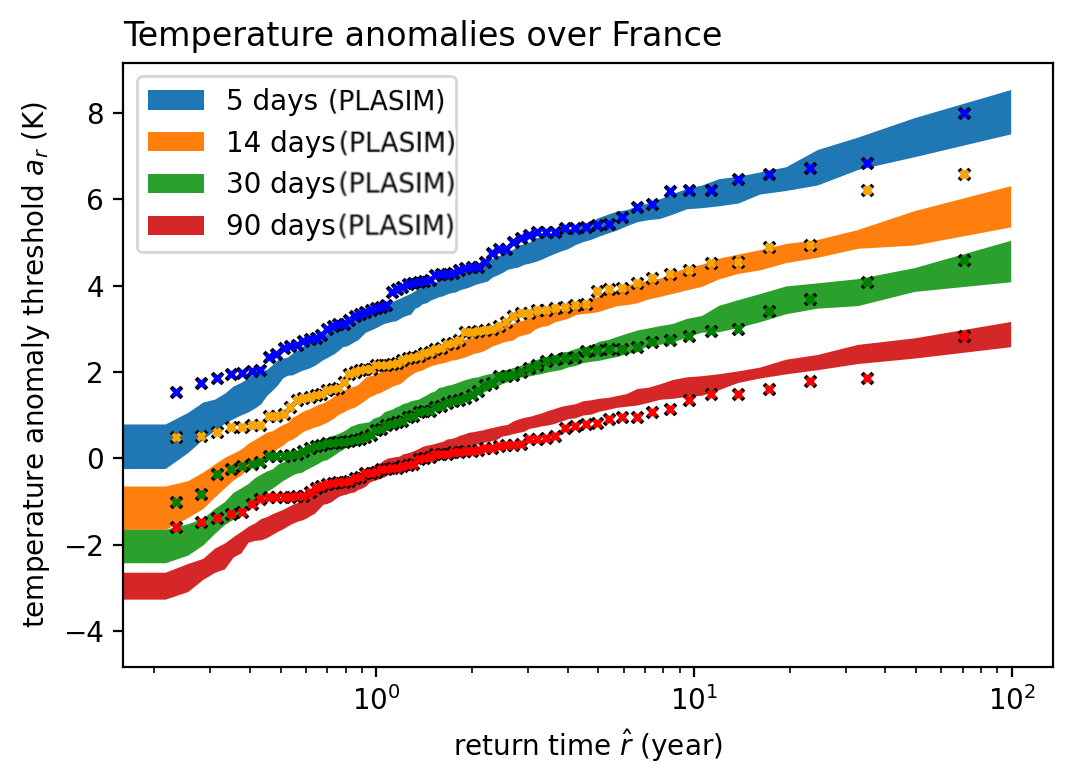}
        \caption{PlaSim France}
        \label{fig:ReturnsPlasimFrance}
    \end{minipage}  
    \hfill
\setcounter{figure}{8}
\setcounter{subfigure}{1}
    \begin{minipage}[b]{0.48\textwidth}
        \includegraphics[width=\linewidth]{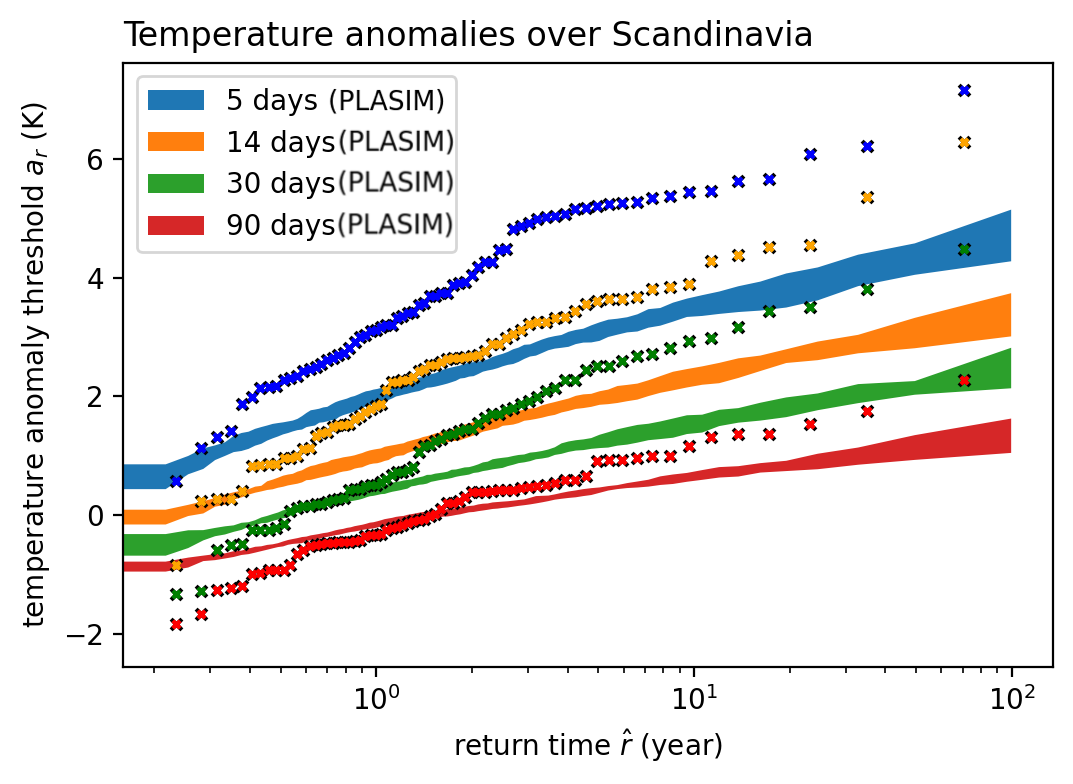}
        \caption{PlaSim Scandinavia}
        \label{fig:ReturnsPlasimScandinavia}
    \end{minipage}
    \hfill
\setcounter{figure}{8}
\setcounter{subfigure}{2}
    \begin{minipage}[b]{0.48\textwidth}
        \includegraphics[width=\linewidth]{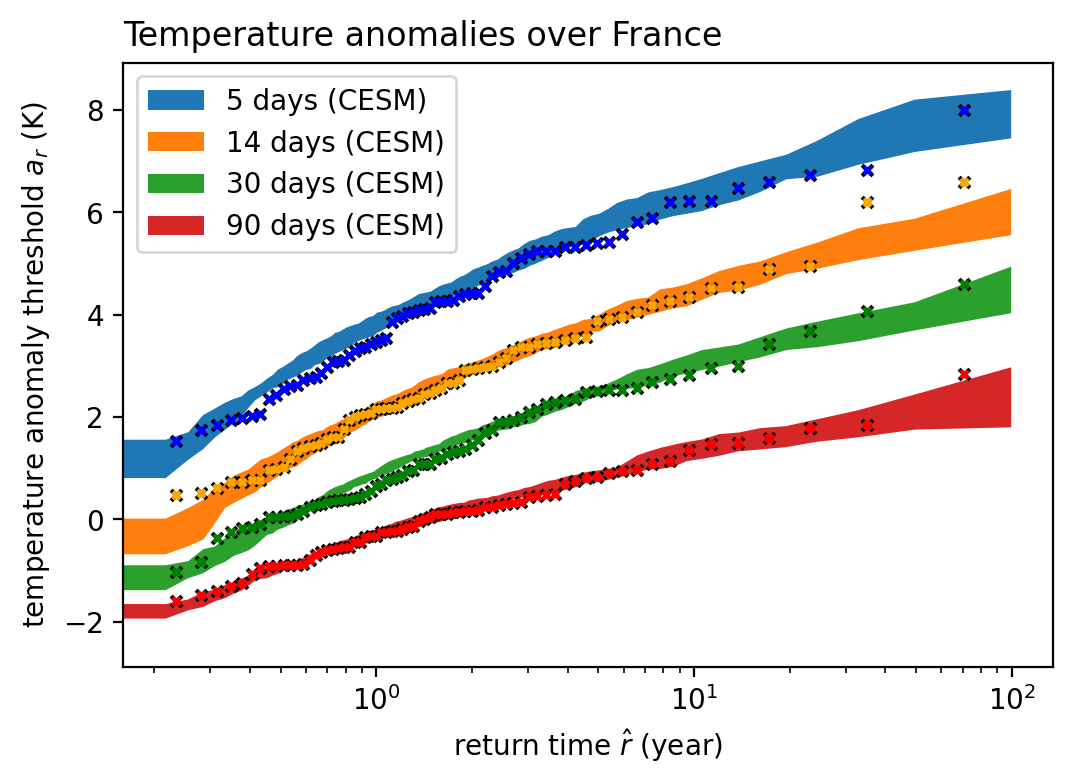}
        \caption{CESM France}
        \label{fig:ReturnsCESMFrance}
    \end{minipage}
    \hfill
\setcounter{figure}{8}
\setcounter{subfigure}{3}
    \begin{minipage}[b]{0.48\textwidth}
        \includegraphics[width=\linewidth]{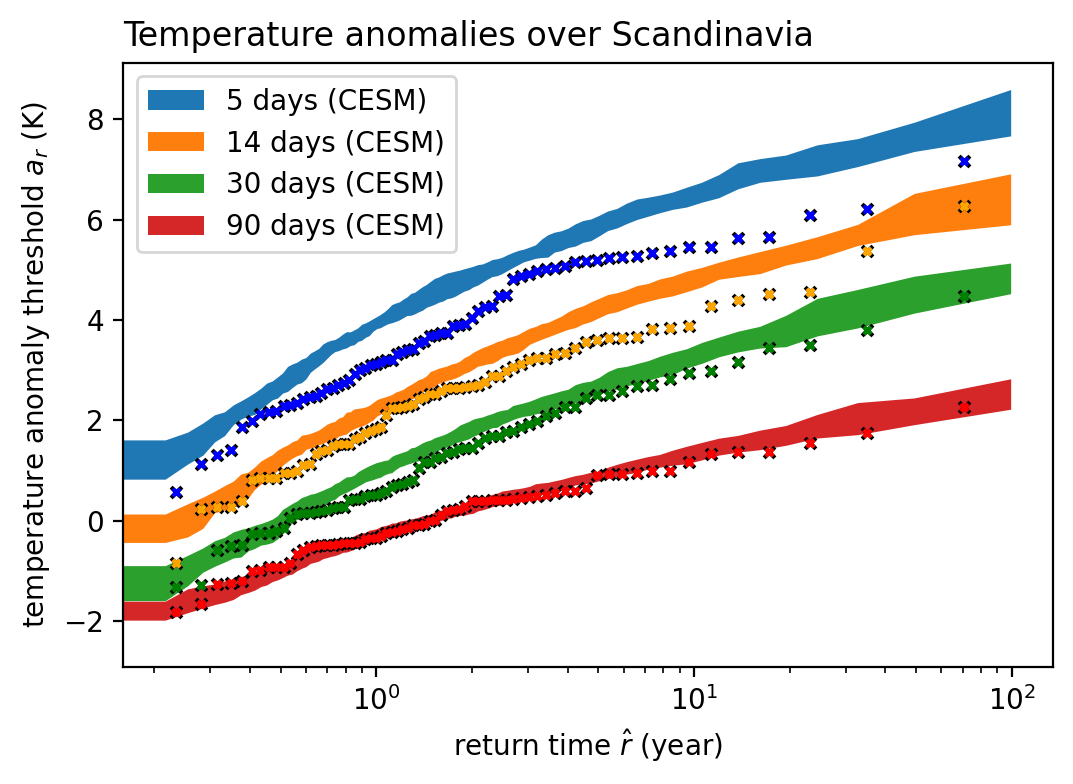}
        \caption{CESM Scandinavia}
        \label{fig:ReturnsCESMScandinavia}
    \end{minipage}

\setcounter{figure}{8}
\setcounter{subfigure}{-1}

    \caption{
Return time plots for heatwaves, summer extremes of two meter temperature anomaly (T2MA), in different regions (left-right) and different models (bottom-top). \textbf{(A)} PlaSim France, \textbf{(B)} PlaSim Scandinavia, \textbf{(C)} CESM France and \textbf{(D)} CESM Scandinavia. The dots correspond to ERA5 data, while the shaded regions to bootstrapping on climate model data, i.e. the data is split into 10 subsets and return times are computed using the approach described in Section~\ref{sec:heatwaves_def_returns}. Then, mean and standard deviation are computed so that the shaded region corresponds to mean plus or minus one standard deviation. The colors and the corresponding duration of heatwaves $T$ are indicated on the legend.
     }
    \label{fig:returns}

\end{subfigure}

Figure~\ref{fig:returns} depicts return time plots computed using equation~\eqref{eq:return_equation}. The left panels correspond to return times for France heatwaves. In particular, Figure~\ref{fig:ReturnsPlasimFrance} shows a remarkable agreement between PlaSim and reanalysis, despite PlaSim being a climate model of intermediate complexity. Long-lasting 90 day heatwaves are fitted even better by CESM (Figure~\ref{fig:ReturnsCESMFrance}). 
If we look at the most extreme 14 day events (orange dots) in ERA5 reanalysis there are two apparent outliers (the most extreme being the European heatwave 2003) compared to the climate models. However, since for these two events we are in the rightmost tail of the distribution where we don't have enough data one could argue that the climate model fits cannot be rejected solely based on these two events. 

For Scandinavia the models do not compare so well with ERA5 reanalysis. For instance, Figure~\ref{fig:ReturnsPlasimScandinavia} demonstrates that PlaSim systematically underestimates the intensity of the extreme events by a large margin and across different values of  $T$, i,e. not only the few extreme outliers. As a result, the 4 year return time has a threshold $a_4 = 1.74 K$ in PlaSim and $a_4 = 3.25 K$ in CESM consistent with the Table~\ref{tab:4yearthresholds}.
Figure~\ref{fig:ReturnsCESMScandinavia} shows that CESM performs much better for Scandinavia. The return times of the events are overestimated, at least for $T<30$ days. In other words, relationship between thresholds and return times is not the same. Nevertheless, it is much closer to reanalysis, for instance for 4 year return time events the predicted threshold is $a_4 = 3.79 K$ (Table~\ref{tab:4yearthresholds}).

\setcounter{figure}{9}
\setcounter{subfigure}{0}
\begin{subfigure}
    \centering
    \begin{minipage}[b]{0.32\textwidth}
        \includegraphics[width=\linewidth]{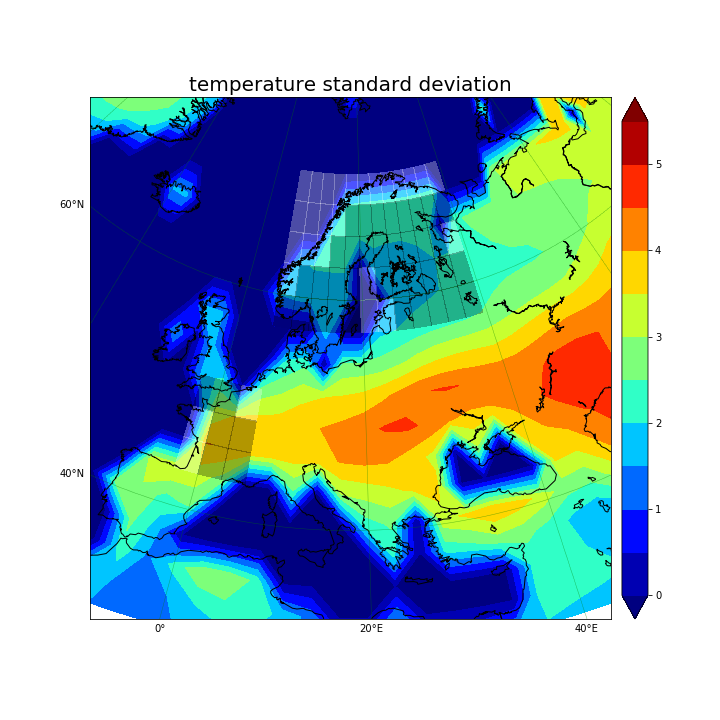}
        \caption{PlaSim}
        \label{fig:stdPlasim}
    \end{minipage}  
    \hfill
\setcounter{figure}{9}
\setcounter{subfigure}{1}
    \begin{minipage}[b]{0.32\textwidth}
        \includegraphics[width=\linewidth]{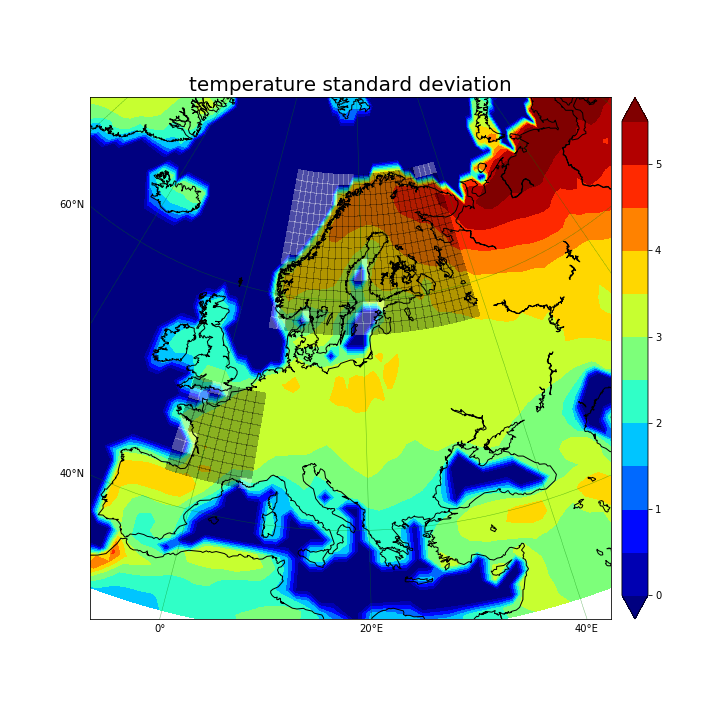}
        \caption{CESM}
        \label{fig:stdCESM}
    \end{minipage}
    \hfill
\setcounter{figure}{9}
\setcounter{subfigure}{2}
    \begin{minipage}[b]{0.32\textwidth}
        \includegraphics[width=\linewidth]{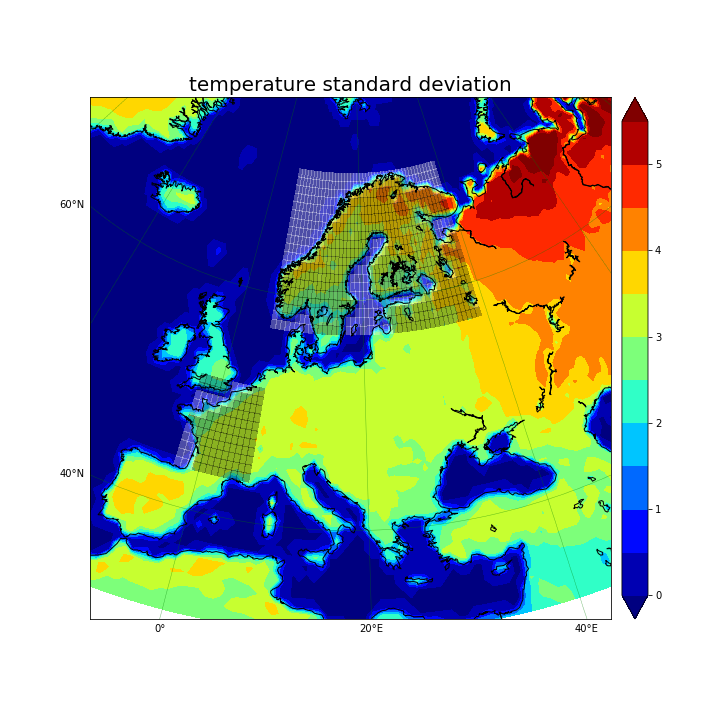}
        \caption{ERA5}
        \label{fig:stdERA5}
    \end{minipage}

\setcounter{figure}{9}
\setcounter{subfigure}{-1}

    \caption{
T2M standard deviation over JJA in different data sets: \textbf{(A)} PlaSim, \textbf{(B)} CESM, \textbf{(C)} Detrended ERA5 reanalysis. Relevant areas (France and Scandinavia) that are used below for regional analysis are shaded.
     }
    \label{fig:local_std}

\end{subfigure}

One may wonder why models have difficulty accurately describing Scandinavia heatwaves. From the theory of single time scale Gaussian stochastic fluctuations, return times depend on the variance and decorrelation time scale of the time series ~\citet{Lestang_2018}. The picture is more complicated when dealing with time averages of signals with multiple time scales. However, generally speaking, return times are expected to scale with the variance of the underlying probability distribution (See Section~\ref{sec:autocovariance}). Figure~\ref{fig:local_std} shows standard deviation of the grid-point 2 meter temperature anomaly (T2MA) PDF taken over JJA period in PlaSim CESM and ERA5. It demonstrates that the models have trouble capturing precisely the variance over the region of north Russia/ Scandinavia, where PlaSim has a strong negative bias while CESM a small positive bias. This goes hand-in-hand with how ERA5 return time plots are underestimated in Figure~\ref{fig:ReturnsPlasimScandinavia} and overestimated in Figure~\ref{fig:ReturnsCESMScandinavia}. In general, according to Figure~\ref{fig:local_std}, PlaSim captures poorly standard deviation in higher latitudes. If we consider other parts of the European continent differences are small and, broadly speaking, CESM compares more favorably to ERA5.


\subsubsection{Autocovariance function}\label{sec:autocovariance}

To have a more complete understanding of the statistics of the tails of the distribution, the standard deviation is not sufficient and the decorrelation properties of the signal must be investigated. We compute the autocovariance function of the T2M time series over France and Scandinavia in PlaSim and CESM and compare them to the corresponding time series in ERA5. The autocovariance function is given by 

\begin{equation}
C(t):=\mathbb{E}[A(X(t)) A(X(0))],
\end{equation}
When the process can be described by a single time scale one expects exponential scaling and the corresponding time scale can be extracted simply as
\begin{equation}\label{eq:integralscale}
    \tau_{c}:=\int_{0}^{+\infty} \frac{{C}(t)}{\mathbb{E}[A^2(X(0))]} \mathrm{d} t
\end{equation}


A similar analysis has been performed in ~\citet{Ragone:2020vs} on 1000 years of data generated with PlaSim in perpetual summer conditions. In that case it was found that time series of area averaged surface temperatures show  two time-scales, the \emph{fast} one of order of 4 days related to synoptic variations and the \emph{slow} one of the order of 30 days, 
which the authors relate to heat capacity of the soil and therefore the soil moisture content.


As stated above, the properties of the autocovariance function provide a contribution to the return time plot. As is known from the theory of stochastic processes~\citet{gardiner1985handbook} a Gaussian process can be generated knowing the parameters of the autocovariance function. We compute it using the PlaSim and CESM  time series of T2M integrated over the area of France and Scandinavia and plot the results on Figure~\ref{fig:autocovariance}. The plots show that the processes involved indeed are well represented by two time scales, which means they can be fitted by two exponential functions. The fit parameters do not precisely match the ones provided by~\citet{Ragone:2020vs} which is not surprising since in that study a heatwave over much larger European region was considered in the perpetual summer regime. Another important parameter to be inferred from Figure~\ref{fig:autocovariance} is the \emph{cross-over scale}, which corresponds to the time at which the slow time scale starts to dominate the decay. 

The presence of two time scales motivates performing exponential fit, using the following nomenclature for its parameters
\begin{equation}\label{doubleexponentfit}
    f = A_1 \exp\left( -\frac{t}{\tau_1} \right) +  A_2 \exp\left( -\frac{t}{\tau_2} \right),
\end{equation}
with $\tau_1 < \tau_2$ the \emph{fast} and \emph{slow} time-scales respectively. The importance of the slow time scale depends on the ratio $A_1/A_2$. The fit is applied to PlaSim/CESM/ERA5 for both France and Scandinavia T2M series and the parameters of the fit are displayed in Table~\ref{tab:autocovariance_parameters}. According to this table {both PlaSim and CESM approximate the first time scale (slope). Indeed, this can be confirmed across all the panels in  Figure~\ref{fig:autoPlasimFrance}. However, Figure~\ref{fig:autoPlasimScandinavia} shows that in Scandinavia PlaSim does not get the variance of the time series right since the blue and the red curve are not matching (also $\tau_1$ is off by approximately factor 2). Another conspicuous feature is the mismatch between the plateau in Figure~\ref{fig:autoPlasimFrance} between PlaSim and ERA5.} These discrepancies translate directly into mismatch between return time plots of PlaSim and ERA5 as displayed in Figure~\ref{fig:ReturnsPlasimScandinavia} already discussed earlier.

\begin{table}
    \begin{center}
     \begin{tabular}{||l c c c c||l c c c c||} 
     \hline
     France & $A_1$ & $A_2$ & $\tau_1$ & $\tau_2$ &
     Scandinavia & $A_1$ & $A_2$ & $\tau_1$ & $\tau_2$ \\ [0.5ex] 
     \hline\hline
     PlaSim & $6.0$ & $1.9$ & $2.1$ & $108$ & PlaSim & $2.1$ & $0.13$ & $2.7$ & $82$ \\ 
     \hline
     CESM & $6.5$ & $0.69$ & $3.4$ & $38.6$ &
     CESM & $7.1$ & $0.28$ & $4.6$ & $80$\\
     \hline
     ERA5 & $5.9$ & $0.45$ & $3.4$ & $108$ &
     ERA5 & $5.0$ & $0.38$ & $4.5$ & $62$\\[1ex]
    \end{tabular}
    \end{center}
    \caption{Exponential parameters for equation~\eqref{doubleexponentfit} fit. The coefficients are given inside the table for each model (rows) and both areas: France on the left and Scandinavia on the right.  }
    \label{tab:autocovariance_parameters}
\end{table}

CESM fits ERA5 quite well in both France and Scandinavia {(see Figures~\ref{fig:autoCESMFrance} and~\ref{fig:autoCESMScandinavia})}, although in the latter case there is an over-estimation of variance discussed earlier in Section~\ref{sec:returns}. Since ERA5 time series is relatively short we cannot make robust comparisons for the second time scale, which tends to be of the order of 30 to 80 days and, as already mentioned, is likely associated with soil-atmosphere interactions. This quasi-plateau identified in {PlaSim on Figure~\ref{fig:autoPlasimFrance}} can be also identified  in CESM and ERA5 {on Figure~\ref{fig:autoCESMFrance}}, however to a lesser degree. This means that France heatwaves in PlaSim have a bias towards stronger temporal persistence than in observations. CESM fits return time for France heatwaves somewhat better than PlaSim (Figures~\ref{fig:ReturnsPlasimFrance} and~\ref{fig:ReturnsCESMFrance}). Thus one may wonder to what extend fitting the second time scale is indispensable for providing a better fit. What if $A_2 = 0$? We will address this question in the next section.

\setcounter{figure}{10}
\setcounter{subfigure}{0}
\begin{subfigure}
    \centering
    \begin{minipage}[b]{0.48\textwidth}
        \includegraphics[width=\linewidth]{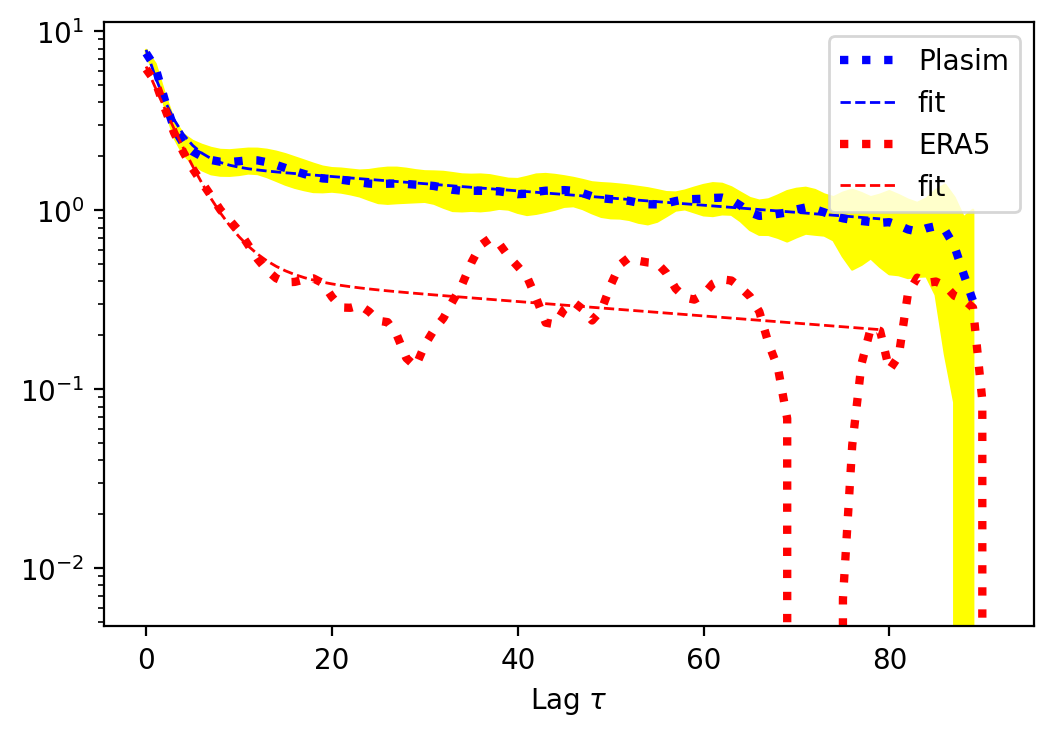}
        \caption{PlaSim France}
        \label{fig:autoPlasimFrance}
    \end{minipage}  
    \hfill
\setcounter{figure}{10}
\setcounter{subfigure}{1}
    \begin{minipage}[b]{0.48\textwidth}
        \includegraphics[width=\linewidth]{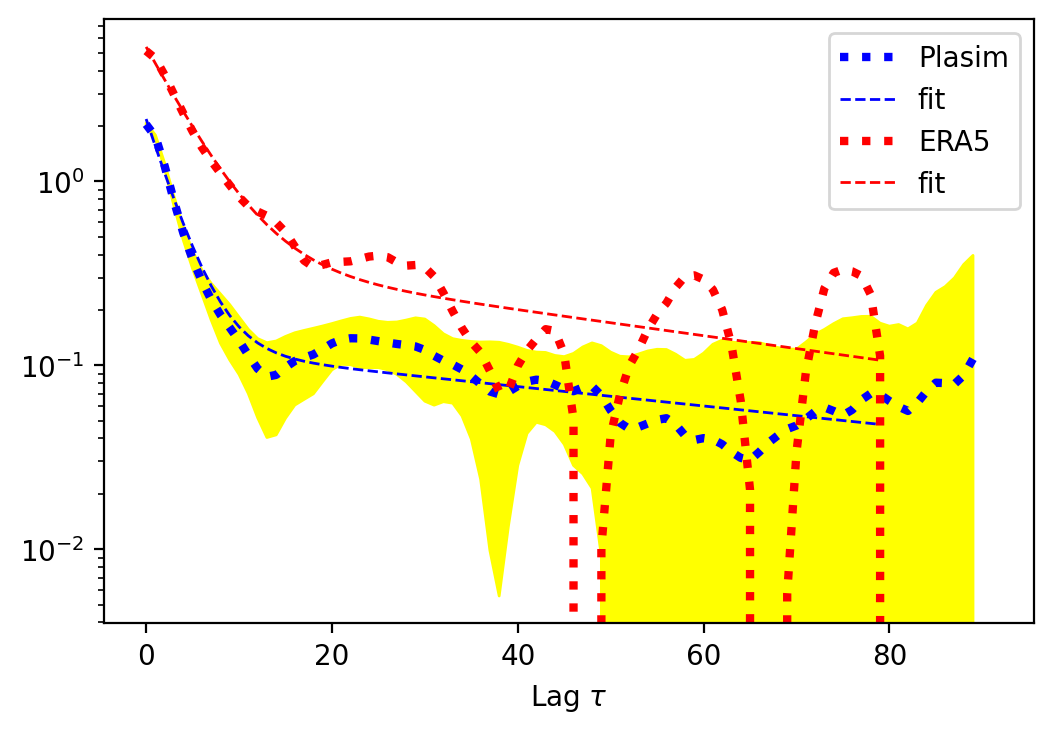}
        \caption{PlaSim Scandinavia}
        \label{fig:autoPlasimScandinavia}
    \end{minipage}
    \hfill
\setcounter{figure}{10}
\setcounter{subfigure}{2}
    \begin{minipage}[b]{0.48\textwidth}
        \includegraphics[width=\linewidth]{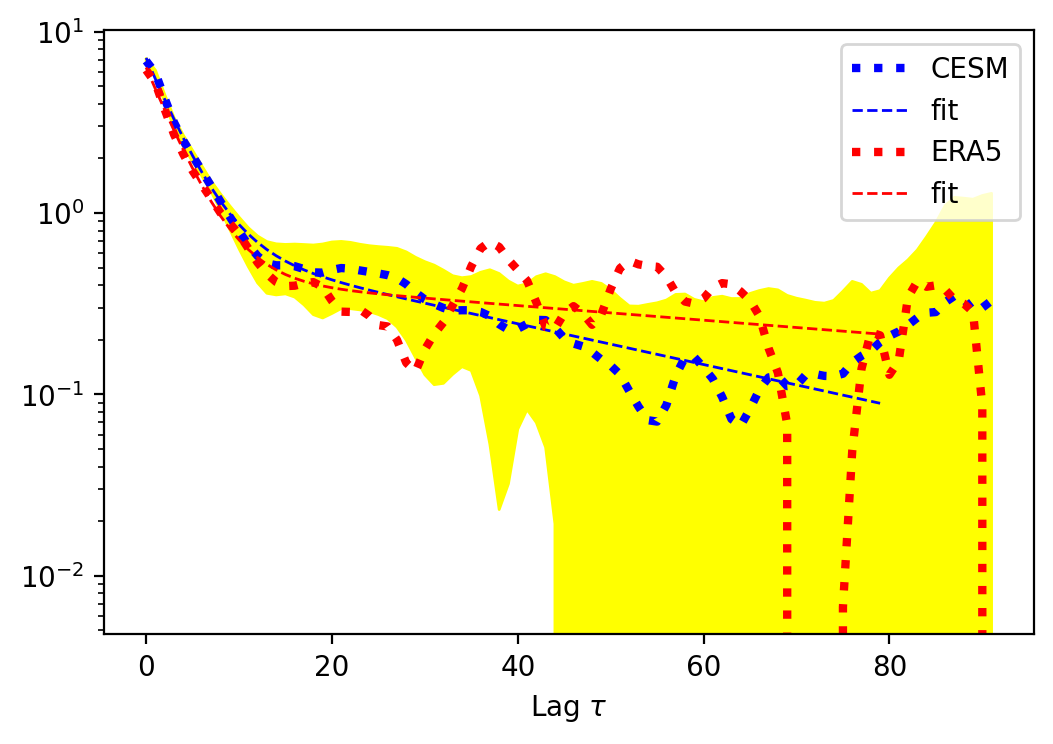}
        \caption{CESM France}
        \label{fig:autoCESMFrance}
    \end{minipage}
    \hfill
\setcounter{figure}{10}
\setcounter{subfigure}{3}
    \begin{minipage}[b]{0.48\textwidth}
        \includegraphics[width=\linewidth]{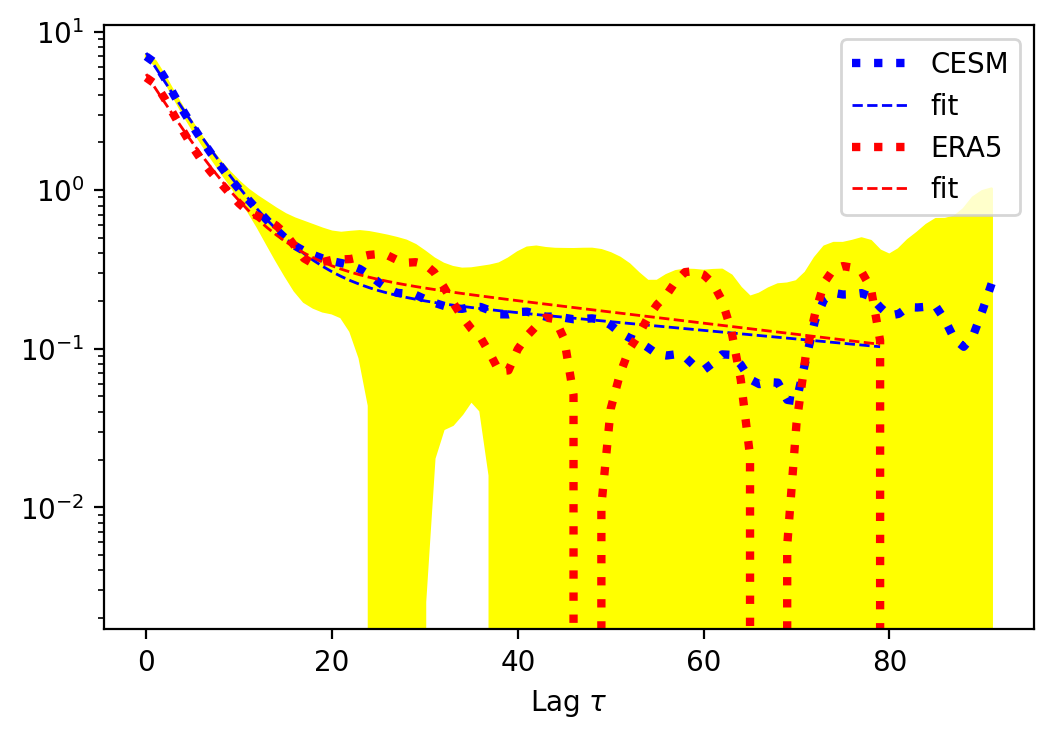}
        \caption{CESM Scandinavia}
        \label{fig:autoCESMScandinavia}
    \end{minipage}

\setcounter{figure}{10}
\setcounter{subfigure}{-1}

    \caption{
Autocovariance function across two regions (left-right) and two models (top-bottom): \textbf{(A)} PlaSim France, \textbf{(B)} PlaSim Scandinavia, \textbf{(C)} CESM France and \textbf{(D)} CESM Scandinavia. The blue dotted curves correspond to the 1000 year-long time series, while the blue dashed lines represent the fit given by equation~2. To provide an estimate of uncertainty we have performed bootstrapping (splitting the time series into 10 trajectories) indicated by the yellow region which represents mean plus or minus one standard deviation. The ERA5 autocovariance function over 71 years is displayed via red dashed line and the underlying time series by red dotted line.
     }
    \label{fig:autocovariance}

\end{subfigure}



\subsubsection{Gaussian process}\label{gauss_procc}

In this Section we compare the statistics of the time series presented above to the synthetic ones generated from  Ornstein-Uhlenbeck (OU) process
\begin{equation}\label{OUequation}
    \mathrm{d} X_{t}^{\alpha,\epsilon}=-\alpha X_{t} \mathrm{~d} t+\sqrt{2\epsilon}\mathrm{d} W_{t}
\end{equation}
which can be interpreted as reformulation of a Langevin equation with correlation time $\tau_c= \alpha^{-1}$~\citet{gardiner1985handbook}. From the corresponding Fokker-Planck equation one finds a stationary probability density $P_s(x)$ that is a Gaussian distribution with standard deviation $\sigma = \sqrt{\epsilon/\alpha}$. 
\begin{equation}
    P_s(x) = \sqrt{\frac{\alpha}{\pi \epsilon}} \exp \left( {-\frac{\alpha x^2}{\epsilon}} \right)
\end{equation}

\begin{equation}
    \langle X(t) X(s) \rangle_s = \frac{\epsilon}{2\alpha} \exp \left( -\alpha |t - s| \right)
\end{equation}
The first passage time conditioned on the stationary measure~\citet{Lestang_2018} can be computed analytically for this model.

However, it is not so simple to compute analytically the autocovariance of the time averaged series. We will therefore adopt an empirical approach following \citet{Herbert2017}. We  simulate the OU process
based on the parameters obtained from the exponential fit on Figure~\ref{fig:autocovariance} and prepare our Gaussian process by a direct sum of the two time series obtained from the OU integrator.
\begin{equation}\label{twotimescaleOU}
    X_t^{\alpha,\epsilon} = X_t^{\alpha_1,\epsilon_1} + X_t^{\alpha_2,\epsilon_2}
\end{equation}
The parameters are chosen as $\alpha_i = 1/\tau_i$ and $\sqrt{2\epsilon_i} = A_i \alpha_i$ (see Table~\ref{tab:autocovariance_parameters}). The synthetic time series of length $10^5$ (``days'') are generated which we then separate into 10 instances of 100 year-long sequences, each of the length  100 ``days'' in order to mimic the summer sequences from the models. The mean and standard deviation is computed based on these 10 instances. 
The resulting return times (equation~\eqref{eq:return_equation}) are plotted on Figure~\ref{fig:returns_gaussian} against the ones from General Circulation Models (GCMs) that were shown on Figure~\ref{fig:returns}. We observe that the match between the Gaussian process and GCMs is relatively good, especially in case of PlaSim over France (Figure~\ref{fig:PlasimFranceDouble}): the model points are within the error bars of the bootstrapped OU simulation.  This is consistent with the fact that the corresponding fluctuations in the upper tail of the distribution are rather Gaussian (see Figure~\ref{fig:distributions}). Both short and long duration heatwaves are captured relatively well (within the error bars), although in some cases the Gaussian process tends to produce less extreme events.

\setcounter{figure}{11}
\setcounter{subfigure}{0}
\begin{subfigure}
    \centering
    \begin{minipage}[t]{0.32\textwidth}
        \includegraphics[width=\linewidth]{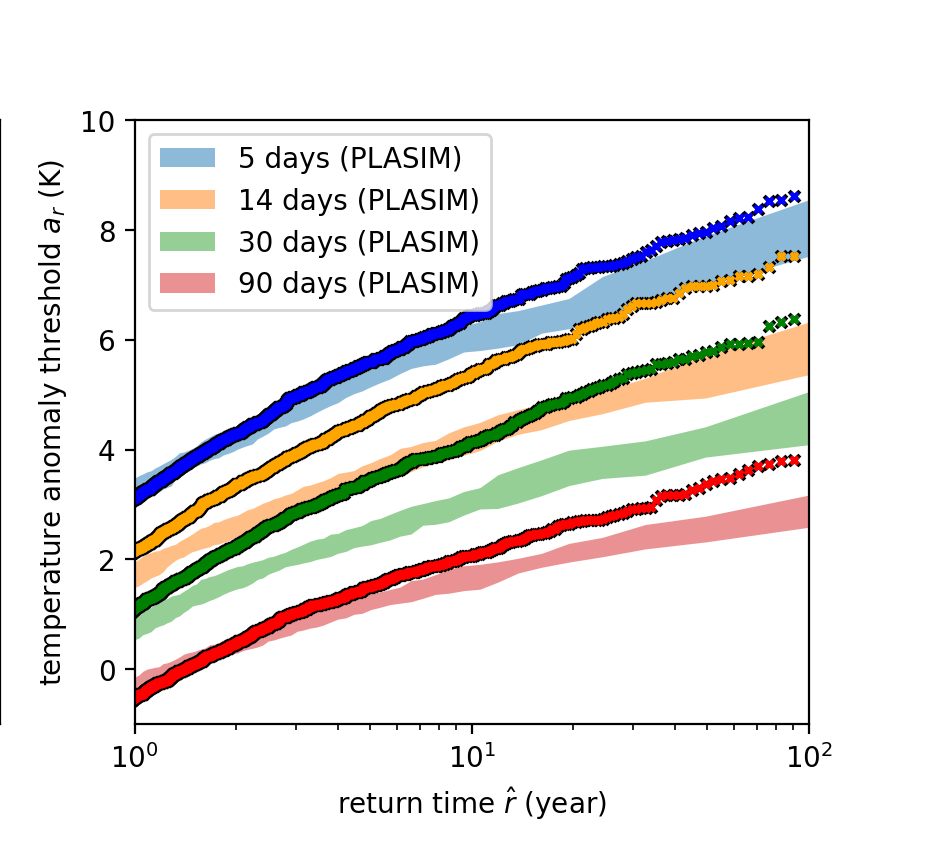}
        \caption{PlaSim France, Single Scale}
        \label{fig:PlasimFranceSingle}
    \end{minipage}  
    \hfill
\setcounter{figure}{11}
\setcounter{subfigure}{1}
    \begin{minipage}[t]{0.32\textwidth}
        \includegraphics[width=\linewidth]{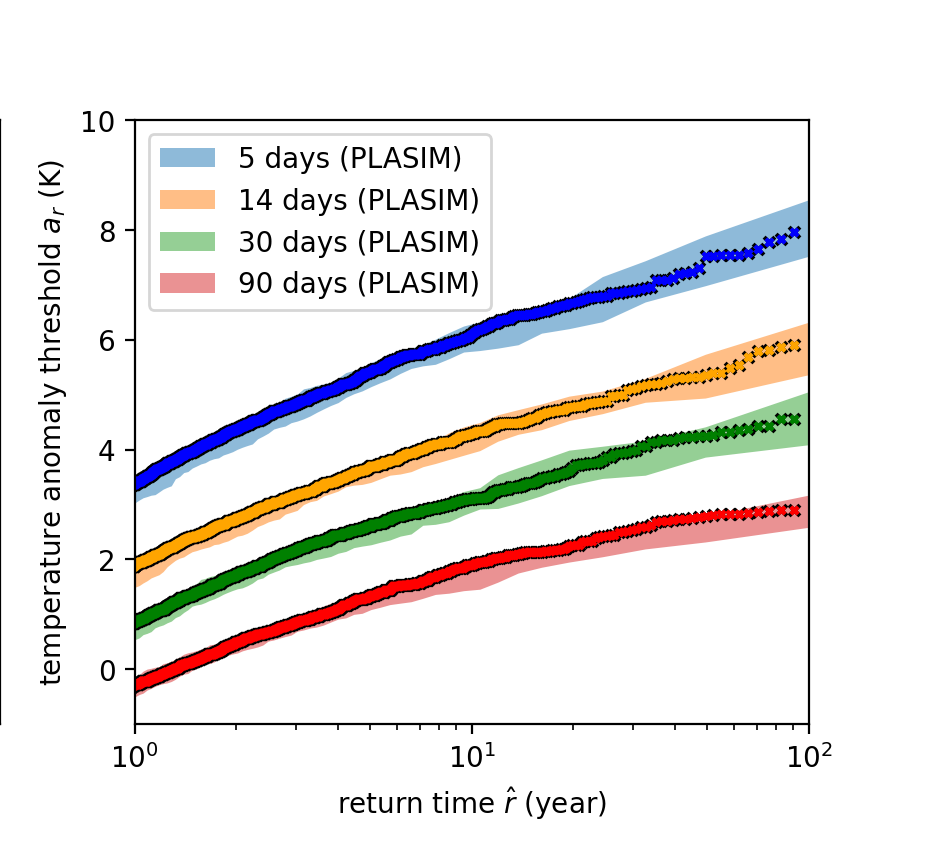}
        \caption{PlaSim France}
        \label{fig:PlasimFranceDouble}
    \end{minipage}
    \hfill
\setcounter{figure}{11}
\setcounter{subfigure}{2}
    \begin{minipage}[t]{0.32\textwidth}
        \includegraphics[width=\linewidth]{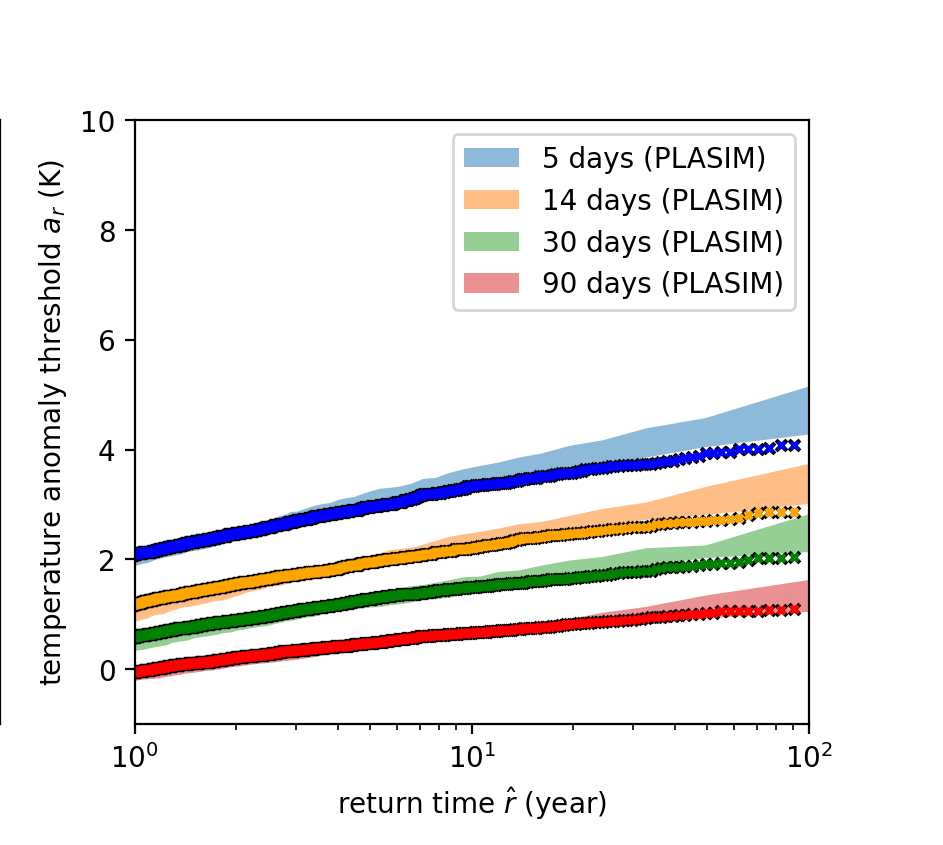}
        \caption{PlaSim Scandinavia}
        \label{fig:PlaSimScandinaviaDouble}
    \end{minipage}

\setcounter{figure}{11}
\setcounter{subfigure}{3}
    \begin{minipage}[t]{0.32\textwidth}
        \includegraphics[width=\linewidth]{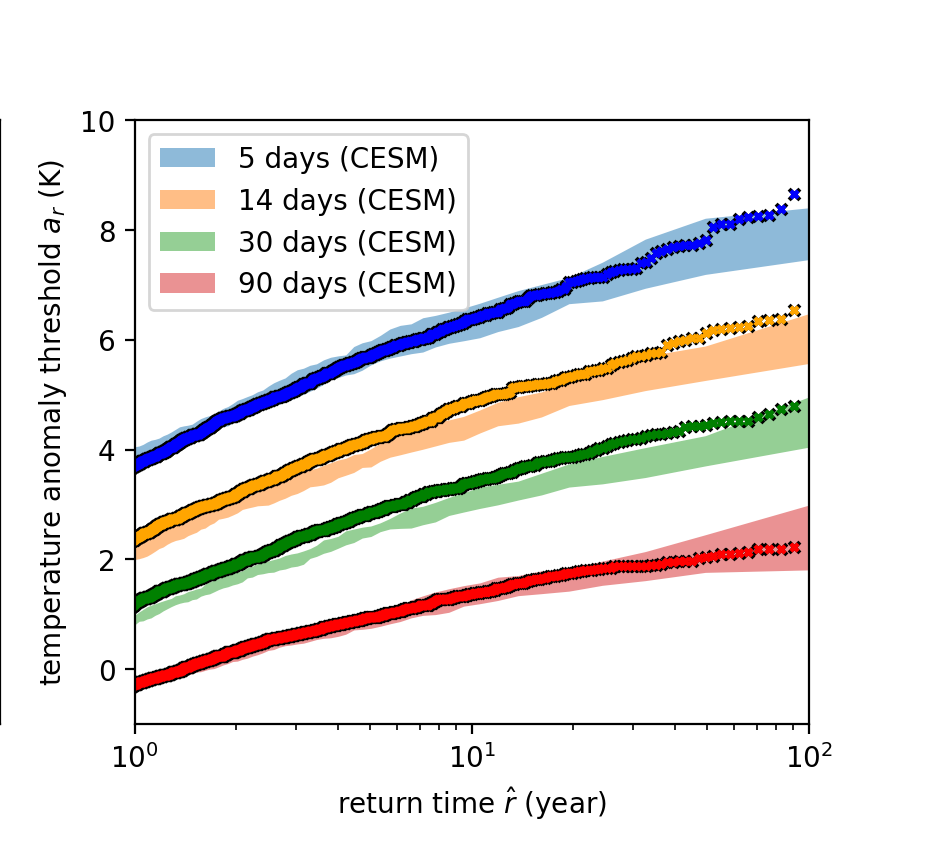}
        \caption{CESM France, single time scale}
        \label{fig:CESMFranceSingle}
    \end{minipage}  
    \hfill
\setcounter{figure}{11}
\setcounter{subfigure}{4}
    \begin{minipage}[t]{0.32\textwidth}
        \includegraphics[width=\linewidth]{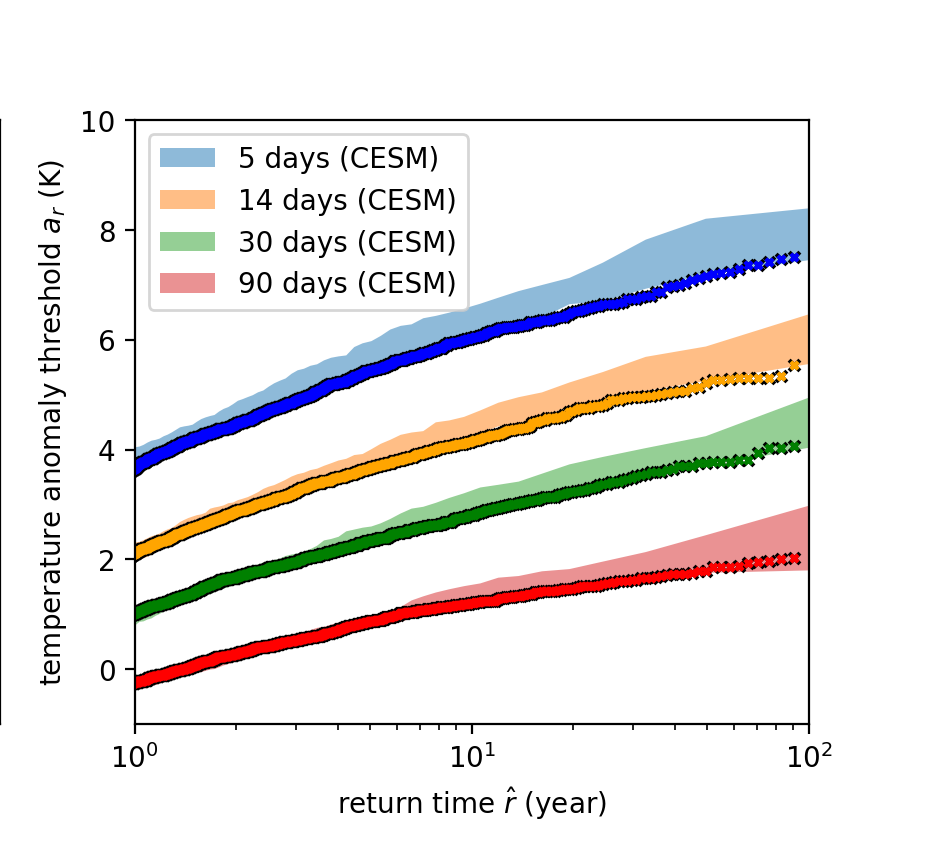}
        \caption{CESM France}
        \label{fig:CESMFranceDouble}
    \end{minipage}
    \hfill
\setcounter{figure}{11}
\setcounter{subfigure}{5}
    \begin{minipage}[t]{0.32\textwidth}
        \includegraphics[width=\linewidth]{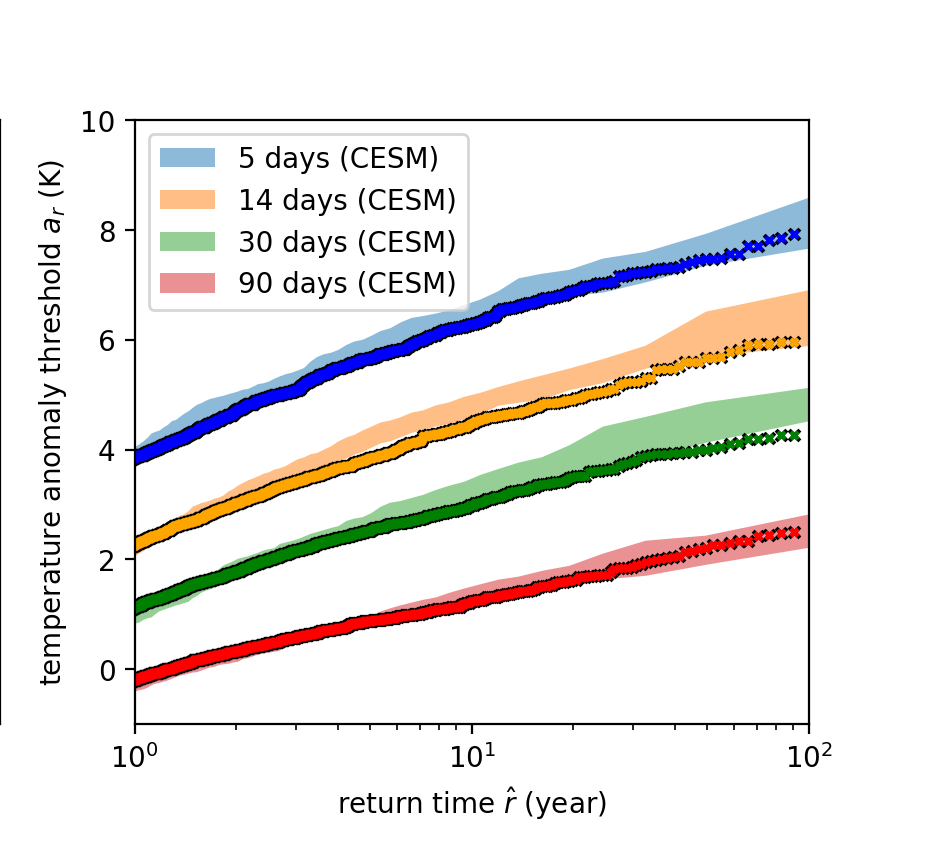}
        \caption{CESM Scandinavia}
        \label{fig:CESMScandinaviaDouble}
    \end{minipage}

\setcounter{figure}{11}
\setcounter{subfigure}{-1}

    \caption{Return time plots for heatwaves in both regions for $T = \{5,14,30,30\}$ days. \textbf{(A)} PlaSim France single time scale, \textbf{(B)} PlaSim France two time scales, \textbf{(C)} PlaSim Scandinavia two time scales, \textbf{(D)} CESM France single time scale, \textbf{(E)} CESM France two time scales and \textbf{(F)} CESM Scandinavia two time scales. The dots correspond to the synthetic time series obtained from OU process~\eqref{OUequation}. (left) single time scale OU process with (top) Fit parameters: $A = 8.101, \tau = 18.6$ days, where the former corresponds to the total variance of the PlaSim data and the latter to the corresponding integral time scale and using the same approach (bottom) Fit parameters: $A = 7.168, \tau = 7$ days. (center, right) two time scale OU process obtained by a direct sum , equation~\eqref{twotimescaleOU}, with the parameters taken from the Figure~\ref{fig:autocovariance}. This figure illustrates that we need two time scales to reproduce well the return time plots. The approach works slightly worse in PlaSim Scandinavia, where likely this is due to the strong kurtosis in the model (See Figure~\ref{fig:distributions}).   }
    \label{fig:returns_gaussian}

\end{subfigure}


Now we are in a position to asses the importance of the second time scale. This is particularly interesting over France where the corresponding coefficient is larger (see Table~\ref{tab:autocovariance_parameters}). To answer this question we only retained the first term in equation~\eqref{doubleexponentfit} and chose the coefficient $A_1 \rightarrow A_1+A_2$ and $A_2\rightarrow 0$ to match the true variance and the integral time scale for $\tau_1 \rightarrow \tau_c$, $\tau_2 \rightarrow 0$, equation~\eqref{eq:integralscale}. The mismatch is quite pronounced for PlaSim (Figure~\ref{fig:PlasimFranceSingle}). In the case of CESM (Figure~\ref{fig:CESMFranceSingle}), on the other hand, the differences are modest but are still noticeable as the extremes start to diverge outside of the error bars. Thus we conclude that the second time scale is less relevant in CESM yet it helps to more accurately capture the return time plot.

\section{Conclusion}

In this paper we presented a spatio-temporal analysis of persistent heatwaves over France and Scandinavia. One of the main findings of this study is that two-week long summer heatwaves are associated with robust teleconnection patterns that involve a strong positive 500 mbar geopotential anomaly (500GPHA) in the region of interest, negative 500GPHA near Greenland and positive  500GPHA in North-East Canada. The patterns are qualitatively consistent across the ERA5 reanalysis and two models of different level of complexity, PlaSim and CESM. 

This quasi-stationary structure appears in Hayashi spectra as a wave-number 3 peak isolated from the general eastward-propagating Rossby wave spectrum. We note that the same wave-number 3 pattern is  identified when a neural network is trained to predict heatwaves using a longer dataset generated with PlaSim ~\citet{miloshevich2022probabilistic} (but with diurnal cycle). Similar patterns and spectral features can be found during the July 2018 heatwave over Scandinavia. In the literature there has been a discussion \citet{Petoukhov2013,petoukhov16,petoukhov18,Kornhuber_2019,Kornhuber20} around the amplification of quasi-stationary Rossby waves with a different range of wavenumbers, between 5 and 8. 

The teleconnection patterns are robust to changes in a period (length) of heatwave as long as it is sufficiently large (larger than several days). The main features (wave-number 3 pattern) are consistent across less extreme (with few year return times) and more extreme (from 4 to 100 year return times). 

We have also shown that the statistical properties of area averaged surface temperature in the two considered regions can be modelled using a two time-scale Ornstein-Ulhenbeck process. The length of the model runs (1000 years) allows us to draw conclusions on the quality of the fits (estimating uncertainty) of  extreme heatwaves with returns as large as 100 years. This allows us to respond positively to the question on whether correctly capturing variance and autocovariance function of the process allows to properly infer the return time curves. The return times of climate models are compared with reanalysis which reveals quite good agreement for heatwaves in France. For the heatwaves in Scandinavia return times in CESM and ERA5 diverge for periods of order 5 to 14 days but are generally well captured for longer events. 

These results demonstrate two things. First, for the regions involved in this study, numerical climate models even of moderate complexity can reproduce the main dynamical and statistical features of persistent heatwaves at {global} scale. In other words, long-lasting, time averaged phenomena are not only more important from the point of view of impacts, but also to a good extent easier to study with numerical models than shorter phenomena. Second, the presence of recurrent large scale teleconnection patterns during long lasting heatwaves over these regions indicates that these events are characterized by a high degree of typicality. This could be potentially very useful in order to identify precursors for the risk of extreme persistent heatwaves, thus improving their predictability, if the dynamical reasons for the formation of these patterns could be properly identified. Recent studies have suggested that typicality of heatwave teleconnection patterns found in reanalysis and CMIP6 models may be compatible with the concept of instanton in large deviation theory ~\citet{galfi2021applications,Galfi21}. While this is a suggestive proposition, we stress that comprehensive quantitative analysis, beyond visual similarity of two-dimensional maps of selected fields, is necessary to support this type of hypothesis.

{In this study we have compared detrended ERA5 reanalysis time series with time series from simulations performed at stationary state, with fixed CO2 concentration and prescribed sea surface temperatures. It would be valuable to examine how teleconnection patterns may change under different greenhouse gas emission scenarios and how these changes could impact the frequency, intensity, and duration of heatwaves. In general, the northward shift in jet streams~\citet{osman21} is expected to contribute to drying regimes in Europe. A substantial increase in high pressure systems over UK in late summers of the second half of the 21st century is also projected ~\citet{rousi21}. This regime is linked with particularly dry conditions in western Europe and may be caused by slow-down of Atlantic Meridional Overturning Circulation (AMOC) according to some studies~\citet{haarsma2015decelerating,duchez2016drivers}. The regime may also originate as a result of circumglobal Rossby wave patterns associated with meandering jetstream~\citet{Kornhuber_2019,Kornhuber20}. Finally, changes in Eurasian snow cover fraction and shrinking sea ice are also significant factors increasing the likelihood of the more persistent European blocking affects and thus affecting European heatwaves in future warming scenarios~\citet{zhang20}.}

{Understanding the effects of future warming on the appearance and behavior of quasi-stationary Rossby waves that induce heatwaves could shed light on the potential risks associated with such events. In order to further investigate the physical mechanisms underlying the teleconnection patterns observed in this study, an analysis of the output of CMIP6 models could be used to pinpoint the interactions between these patterns and other modes of climate variability, such as North Atlantic Oscillation (NAO), considering that coupled models seem to improve the representation of teleconnection patterns~\citet{rousi21}.}

\section*{Code availability statement}

The coding resources for this work, such as the python and jupyter notebook files, are available on a GitHub page \href{https://github.com/georgemilosh/Climate-Learning}{https://github.com/georgemilosh/Climate-Learning} and is part of a larger project at LSCE/IPSL ENS de Lyon with multiple collaborators working on rare event algorithms.

\section*{Acknowledgement}

This work was supported by the ANR grant SAMPRACE, project ANR-20-CE01-0008-01 (F. Bouchet). This work has received funding through the ACADEMICS grant of the IDEXLYON, project of the Universit\'e de Lyon, PIA operated by ANR-16-IDEX-0005. We acknowledge CBP IT test platform (ENS de Lyon, France) for ML facilities and GPU devices, operating the SIDUS solution~\citet{SIDUS}. This work was granted access to the HPC resources of CINES under the DARI allocations A0050110575,  A0070110575, A0090110575 and A0110110575 made by GENCI.  We acknowledge the help of Alessandro Lovo in maintaining the GitHub page.


\bibliographystyle{Frontiers-Harvard}
\bibliography{test}
\end{document}